\pdfoutput=1
\documentclass[11pt,twoside,a4paper,cmspaper,final,collab]{cms-tdr}
\def\svnVersion{ea16eec}\def\svnDate{2022/05/24}\def\cmsCernNoTag{CERN-EP-2021-189}\def\cmsCernDate{\today}\def\cmsMessage{Published in the Journal of High Energy Physics as \href{http://dx.doi.org/10.1007/JHEP06(2022)012}{\doi{10.1007/JHEP06(2022)012}.}}
\begin{document}\cmsNoteHeader{HIG-20-006}

\providecommand{\cmsTable}[1]{\resizebox{\textwidth}{!}{#1}}

\newcommand{\muqqh} {\ensuremath{\mu_{\Pq\Pq\PH}}\xspace} 
\newcommand{\muggh} {\ensuremath{\mu_{\Pg\Pg\PH}}\xspace} 

\newcommand{\taum}{\ensuremath{\PGt_{\Pgm}}\xspace}
\newcommand{\taue}{\ensuremath{\PGt_{\Pe}}\xspace}
\newcommand{\taul}{\ensuremath{\PGt_{\ell}}\xspace}
\newcommand{\Paone} {\ensuremath{\mathrm{a_{1}}}\xspace}

\newcommand{\ggH} {\ensuremath{\Pg\Pg\PH}\xspace} 
\newcommand{\VH} {\ensuremath{\PV\PH}\xspace} 
\newcommand{\HTT} {\ensuremath{\PH\to\PGt\PGt}\xspace} 
\newcommand{\ZTT} {\ensuremath{\PZ\to\PGt\PGt}\xspace} 
\newcommand{\rhotopipiz} {\ensuremath{\PGr^{\pm}\to\Pgppm\Pgpz}\xspace} 
\newcommand{\Wjets} {\ensuremath{\PW+\text{jets}}\xspace} 
\newcommand{\Zjets} {\ensuremath{\PZ+\text{jets}}\xspace} 
\newcommand{\Zll} {\ensuremath{\PZ\to\ell\ell}\xspace} 
\newcommand{\Zmm} {\ensuremath{\PZ\to\Pgm\Pgm}\xspace} 
\newcommand{\Zee} {\ensuremath{\PZ\to\Pe\Pe}\xspace} 
\newcommand{\Zmt} {\ensuremath{\PZ\to\mutau}\xspace} 
\newcommand{\mutotauh} {\ensuremath{\Pgm\to\tauh}\xspace} 
\newcommand{\etotauh} {\ensuremath{\Pe\to\tauh}\xspace} 
\newcommand{\tautau} {\ensuremath{\PGt\PGt}\xspace} 

\newcommand{\phicp} {\ensuremath{\phi_{\CP}}\xspace} 
\newcommand{\phitt} {\ensuremath{\alpha^{\PH\tau\tau}}\xspace} 
\newcommand{\fCPtt} {\ensuremath{f_{\CP}^{\PH\PGt\PGt}}\xspace} 

\newcommand{\ltau} {\ensuremath{\taul\tauh}\xspace}
\newcommand{\Ltau} {\ensuremath{\ell+\tauh}\xspace}
\newcommand{\etau} {\ensuremath{\taue\tauh}\xspace}
\newcommand{\mutau} {\ensuremath{\tau_{\Pgm}\tauh}\xspace}
\newcommand{\tauhtauh} {\ensuremath{\tauh\tauh}\xspace}

\newcommand{\mtt}{\ensuremath{m_{\PGt\PGt}}\xspace}
\newcommand{\Deltaeta}{\ensuremath{\Delta\eta}\xspace}
\newcommand{\Deltaphi}{\ensuremath{\Delta\phi}\xspace}
\newcommand{\Deltaphijj}{\ensuremath{\Delta\phi_{\mathrm{jj}}}\xspace}
\newcommand{\dz}{\ensuremath{d_{\mathrm{z}}}\xspace}
\newcommand{\dxy}{\ensuremath{d_{\mathrm{xy}}}\xspace}
\newcommand{\PP}{\ensuremath{\Pp\Pp}\xspace}
\newcommand{\HPS} {{\text{Hadron-Plus-Strips}}\xspace}
\newcommand{\TAUSPINNER} {{\textsc{tauspinner}}\xspace}
\newcommand{\SVFIT} {{\textsc{Svfit}}\xspace}
\newcommand{\TOPPLUSPLUS} {{\textsc{Top++v2.0}}\xspace}
\newcommand{\CP} {{\textit{CP}}\xspace}
\newcommand{\jettotauh}{\ensuremath{\text{jet}\to\tauh}\xspace}
\newcommand{\totauh}{\ensuremath{\to\tauh}\xspace}
\newcommand{\ltotauh}{\ensuremath{\ell\to\tauh}\xspace}

\newcommand{\Yuktau} {\ensuremath{\PH\PGt\PGt}\xspace}
\newcommand{\Yukttbar} {\ensuremath{\PH\mathrm{tt}}\xspace}
\newcommand{\PaoneONEP} {\ensuremath{\mathrm{a_{1}^{1pr}}}\xspace}
\newcommand{\PaoneTHREEP} {\ensuremath{\mathrm{a_{1}^{3pr}}}\xspace}
\newcommand{\THREEPPIZ} {\ensuremath{\Pgppm\PGpmp\Pgppm\Pgpz}\xspace}
\newcommand{\hpm} {\ensuremath{\Ph^{\pm}}\xspace}
\newcommand{\PGrzero} {\ensuremath{\PGr^0}\xspace}
\newcommand{\aoneaone} {\ensuremath{\PaoneTHREEP\PaoneTHREEP}\xspace} 
\newcommand{\pipi} {\ensuremath{\Pgp\Pgp}\xspace} 
\newcommand{\mupi} {\ensuremath{\Pgm\Pgp}\xspace} 
\newcommand{\epi} {\ensuremath{\Pe\Pgp}\xspace} 
\newcommand{\rhorho} {\ensuremath{\PGr\PGr}\xspace} 
\newcommand{\murho} {\ensuremath{\Pgm\PGr}\xspace} 
\newcommand{\erho} {\ensuremath{\Pe\PGr}\xspace} 
\newcommand{\pirho} {\ensuremath{\Pgp\PGr}\xspace} 

\newcommand{\kappattilde} {\ensuremath{\widetilde{\kappa}_\tau}\xspace}
\newcommand{\kappat} {\ensuremath{\kappa_\tau}\xspace}
\newcommand{\muvec} {\ensuremath{\vec{\mu}}\xspace}
\newcommand{\alphapi} {\ensuremath{\alpha^{\pi}_{-}}\xspace}
\newcommand{\alpharho} {\ensuremath{\alpha^{\rho}_{-}}\xspace}
\newcommand{\phistar} {\ensuremath{\phi^*}\xspace}
\newcommand{\Ostar} {\ensuremath{O^*}\xspace}
\newcommand{\thetaGJ} {\ensuremath{\theta_{\mathrm{GJ}}}\xspace}
\newcommand{\thetaGJmax} {\ensuremath{\theta_{\mathrm{GJ}}^{\text{max}}}\xspace}
\newcommand{\mtau} {\ensuremath{m_{\PGt}}\xspace}
\newcommand{\maone} {\ensuremath{m_{\Paone}}\xspace}
\newcommand{\paone} {\ensuremath{\vec{p}_{\Paone}}\xspace}
\newcommand{\ytau} {\ensuremath{y^{\PGt}}\xspace}
\newcommand{\lambdahat} {\ensuremath{\hat{\lambda}}\xspace}
\newcommand{\lambdahatZMFplus} {\ensuremath{\lambdahat^{ZMF+}_{\bot}}\xspace}
\newcommand{\lambdahatZMFminus} {\ensuremath{\lambdahat^{ZMF-}_{\bot}}\xspace}
\newcommand{\lambdaZMF} {\ensuremath{\lambda^{ZMF}}\xspace}
\newcommand{\qZMF} {\ensuremath{q^{ZMF\pm}}\xspace}
\newcommand{\phiZMF} {\ensuremath{\phi^{ZMF}}\xspace}
\newcommand{\OZMF} {\ensuremath{O^{ZMF}}\xspace}
\newcommand{\vecjpm} {\ensuremath{\vec{j}^\pm}\xspace}
\newcommand{\Epi} {\ensuremath{E_{\Pgppm}}\xspace}
\newcommand{\Epizero} {\ensuremath{E_{\Pgpz}}\xspace}
\newcommand{\ypm} {\ensuremath{y^{\PGtpm}}\xspace}
\newcommand{\vech} {\ensuremath{\vec{h}}\xspace}
\newcommand{\veck} {\ensuremath{\vec{k}}\xspace}
\newcommand{\vecn} {\ensuremath{\vec{n}}\xspace}
\newcommand{\vectheta} {\ensuremath{\vec{\theta}}\xspace}
\newcommand{\zhat} {\ensuremath{\hat{z}}\xspace}
\newcommand{\phat} {\ensuremath{\hat{p}}\xspace}
\newcommand{\jhat} {\ensuremath{\hat{j}}\xspace}

\newcommand{\ptmu} {\ensuremath{\pt^{\Pgm}}\xspace}
\newcommand{\pttauh} {\ensuremath{\pt^{\tauh}}\xspace}
\newcommand{\pte} {\ensuremath{\pt^{\Pe}}\xspace}
\newcommand{\SIP} {\ensuremath{S_{\text{IP}}}\xspace}
\newcommand{\eIso} {\ensuremath{I^{\Pe}}\xspace}
\newcommand{\mIso} {\ensuremath{I^{\Pgm}}\xspace}
\newcommand{\FF} {\ensuremath{F_{\text{F}}}\xspace} 
\newcommand{\likeli} {\ensuremath{L(\lumi,\muvec,\phitt,\vec{\theta})}\xspace} 
\newcommand{\nij} {\ensuremath{n_{\text{i,j}}}\xspace} 
\newcommand{\Bij} {\ensuremath{B_{\text{i,j}}(\vectheta)}\xspace} 
\newcommand{\Aij} {\ensuremath{\vec{A}_{\text{i,j}}(\vectheta,\phitt)}\xspace} 
\newcommand{\Cm} {\ensuremath{C_{\mathrm{m}} (\vectheta)}\xspace} 
\newcommand{\DeltaLL} {\ensuremath{-2 \Delta \ln L}\xspace} 
\newcommand{\AveAsym} {\ensuremath{A\:S/(S+B)}\xspace} 
\newcommand{\CPeven} {\ensuremath{\CP^{\text{even}}}\xspace} 
\newcommand{\CPodd} {\ensuremath{\CP^{\text{odd}}}\xspace} 

\cmsNoteHeader{HIG-20-006} 
\title{Analysis of the \texorpdfstring{\CP}{CP} structure of the Yukawa coupling between the Higgs boson and \texorpdfstring{\PGt}{tau} leptons in proton-proton collisions at \texorpdfstring{$\sqrt{s}=13\TeV$}{sqrt(s) = 13 TeV}}

\date{\today}

\abstract{
The first measurement of the \CP structure of the Yukawa coupling between the Higgs boson and \PGt leptons is presented. The measurement is based on data collected in proton-proton collisions at $\sqrt{s}=13\TeV$ by the CMS detector at the LHC, corresponding to an integrated luminosity of 137\fbinv. The analysis uses the angular correlation between the decay planes of \PGt leptons produced in Higgs boson decays. The effective mixing angle between \CP-even and \CP-odd \PGt Yukawa couplings is found to be $-1\pm 19^{\circ}$, compared to an expected value of $0\pm 21^{\circ}$ at the 68.3\% confidence level. The data disfavour the pure \CP-odd scenario at 3.0 standard deviations. The results are compatible with predictions for the standard model Higgs boson.
}

\hypersetup{
pdfauthor={CMS Collaboration},
pdftitle={Analysis of the CP structure of the Yukawa coupling between the Higgs boson and tau leptons in proton-proton collisions at a centre-of-mass energy of 13 TeV},
pdfsubject={CMS},
pdfkeywords={CMS,  Higgs, taus}}

\maketitle

\section{Introduction}\label{sec:Introduction}

In the standard model (SM), the electroweak symmetry breaking is postulated via the Brout--Englert--Higgs mechanism~\cite{PhysRevLett.13.321, Higgs:1964ia,
PhysRevLett.13.508, PhysRevLett.13.585, Higgs:1966ev, Kibble:1967sv}. This mechanism predicts the existence of a scalar particle, the Higgs boson (\PH). 
A particle compatible with this boson was discovered 
by the ATLAS~\cite{Aad_2012} and CMS~\cite{Chatrchyan_2012, Chatrchyan:2013lba} Collaborations at the LHC using proton-proton (\PP) collision data collected 
in 2011 and 2012 at centre-of-mass energies of 7 and 8\TeV, respectively.
 Since 2012, the couplings of the Higgs boson to heavy quarks, leptons, and gauge bosons have been measured, including the coupling to \PGt leptons~\cite{Khachatryan:2016vau,Sirunyan_2018,Aaboud_2019}, and the most recent measurement of its mass value yields $125.38\pm0.14\GeV$~\cite{Sirunyan:2020xwk}.

The SM \PH is even under charge-parity (\CP) inversion. 
A sizeable deviation from a pure \CP-even interaction of the \PH with any of the SM particles would be a direct indication of physics beyond the SM.
Therefore, the \CP structure of the couplings of the \PH is of paramount interest.
The CMS and ATLAS Collaborations have studied the couplings of the \PH to vector gauge bosons, including tests of \CP violation~\cite{Chatrchyan:2012jja,Chatrchyan:2013mxa,Khachatryan:2014kca,Khachatryan:2015mma,Khachatryan:2016tnr,Sirunyan:2017tqd,Sirunyan_2019,	CMS:2021nnc, Aad:2013xqa,Aad:2015mxa,Aad:2016nal,Aaboud:2017oem,Aaboud:2017vzb,Aaboud:2018xdt}. 
These studies excluded pure pseudoscalar (\CP-odd) interactions of the \PH with the \PW and \PZ bosons (referred to collectively as \PV bosons).

There are strong theoretical motivations to search for \CP-violating effects in couplings of the \PH to fermions rather than \PV bosons. 
In couplings to \PV bosons, \CP-odd contributions enter via nonrenormalisable higher-order operators that are suppressed by powers of
 $1/\Lambda^{2}$~\cite{Zhang_2011,Harnik:2013aja,Ghosh_2019}, where $\Lambda$ is the scale of the physics beyond the SM in an effective field theory.
Therefore, these are expected to only yield a small contribution to the coupling. 
 A renormalisable \CP-violating  Higgs-to-fermion coupling may occur at tree level. The \PGt lepton and top quark Yukawa couplings, \Yuktau and \Yukttbar, respectively, are
 therefore the optimal couplings for \CP studies in \PP collisions~\cite{Gritsan_2016}, and measurements of these two couplings are complementary.
Recently, both the CMS~\cite{CMSTOPCP2020} and ATLAS~\cite{ATLASTOPCP2020} Collaborations presented first measurements of the \CP
structure of the \PH coupling to top quarks. The CMS result rejects the purely \CP-odd hypothesis with a significance of
3.2 standard deviations, $\sigma$, while the ATLAS analysis rejects it with a significance of $3.9\sigma$. 
The CMS and ATLAS Collaborations have also studied the \CP-nature of the \PH interaction with gluons~\cite{CMS:2021nnc,ATLAS:2021pkb} which was found to be consistent with the SM expectation, albeit with limited sensitivity. 
Such studies may also be interpreted in terms of the \PH coupling to top quarks, under the assumption that the interaction is mediated predominantly via top quark loops.

The \CP-properties of the \HTT process is commonly described in terms of an effective mixing angle \phitt, which is virtually equal to $0^{\circ}$ in the SM.
The measurement of a nonzero \phitt would therefore directly contradict the SM predictions, and have implications for beyond the SM physics models, such as two-Higgs-doublet models~\cite{Fontes:2015mea}, including supersymmetry.
For example, in the minimal supersymmetric model \CP violation in the Higgs-to-fermion couplings is expected to be small and therefore the measurement of a sizeable mixing angle would disfavour such scenarios. In contrast, in the next-to-minimal supersymmetric model, \phitt
can be as large as $27^{\circ}$~\cite{King_2015}.
Feasibility studies have indicated that the LHC experiments can measure \phitt to a precision of about 5--$10^{\circ}$ with 3\abinv of data~\cite{Harnik:2013aja,Berge2014}. 

In this paper we present the first measurement of the \CP structure of the \PH coupling to \PGt leptons.
This analysis uses the \PP data sets collected by the CMS detector at $\sqrt{s}=13\TeV$ in 2016, 2017, and 2018.
These correspond to integrated luminosities of 35.9, 41.5, and 59.7\fbinv, respectively.
This analysis targets the most sensitive \tauhtauh, \mutau and \etau decay channels, where a \PGt lepton decaying to hadrons is denoted as \tauh, and a \PGt lepton decaying to a muon or an electron as \taum or \taue (or collectively as \taul), respectively. 
The decays into light leptons are accompanied by two neutrinos, while the hadronic modes involve one neutrino. These particles are not directly detected but result in a transverse momentum imbalance which can be used to partially constrain the \tautau system.
In total, this analysis covers about 70\% of all possible \PGt lepton pair decay modes.
Table~\ref{tab:AnalysisStrategy_TauDecayModes} summarises the \PGt lepton decay modes used in this analysis, their branching fractions, and the shorthand symbols that we use to denote them in the rest of this paper. 
The charged hadrons are denoted by the symbol \hpm, which consist mainly of charged pions but include a smaller contribution from charged kaons. Throughout this paper we will assume that all \hpm are charged pions (\Pgppm) since the CMS detector is not able to distinguish between different types of \hpm.

\begin{table}[hbtp]
\centering
\topcaption{Decay modes of \PGt leptons used in this analysis and their branching fractions $\mathcal{B}$~\cite{Zyla:2020zbs}. Where appropriate, we indicate the known intermediate resonances. The last row gives the shorthand notation for the decays used throughout this paper. \label{tab:AnalysisStrategy_TauDecayModes}}
\begin{tabular}{cccccccc}
\hline
Mode & $\Pepm\nu\nu$ & $\PGmpm\nu\nu$ & $\hpm\nu$ & $\hpm\Pgpz\nu$ & $\hpm\Pgpz\Pgpz\nu$  & $\hpm\Ph^{\mp}\hpm\nu$ \\ \hline
Type &\taue        &\taum           &\tauh   &\tauh  & \tauh  &\tauh \\
$\mathcal{B}(\%)$& 17.8 & 17.4 & 11.5 & 25.9 & 9.5 & 9.8 \\
Resonance & \NA & \NA & \NA & \Pgr & \Pai  & \Pai \\
Symbol &\Pe        &\Pgm           &\Pgp   &\PGr  & \PaoneONEP  &\PaoneTHREEP \\ \hline
\end{tabular}
\end{table}

This paper is organised as follows. 
The parameterisation of the \CP properties of the \PGt Yukawa coupling is discussed in Section~\ref{sec:pheno}.
In Section~\ref{sec:CMSdetector} the experimental setup is outlined, and this is followed by a discussion of the data sets and simulated samples in Section~\ref{sec:datasimulatedsamples}. Subsequently, the event reconstruction is presented in Section~\ref{sec:eventreconstruction}. 
Thereafter, in Section~\ref{sec:decayplanereco} the \CP-sensitive observables used to extract the results are outlined. In Section~\ref{sec:eventselection} the event selection is presented.
The estimation of the backgrounds is discussed in Section~\ref{sec:backgroundestimation}. 
The techniques used to distinguish the signal from the background events are outlined in Section~\ref{sec:eventcategorisation}. In Section~\ref{sec:unrolled} various distributions that are used to extract the results are displayed and discussed.
In Section~\ref{sec:systematicuncertainties} the systematic uncertainties are presented. The results are discussed in Section~\ref{sec:results}, and a summary of the analysis is given in Section~\ref{sec:summary}.

\section{Parametrisation of the \texorpdfstring{\CP}{CP} properties of the \texorpdfstring{\PGt}{tau} Yukawa coupling}\label{sec:pheno}

We parameterise the Lagrangian for the \PGt Yukawa coupling in terms of the coupling strength modifiers \kappat and \kappattilde that parameterise the \CP-even and \CP-odd contributions, respectively~\cite{Gritsan_2016}: 
\begin{linenomath}
\begin{equation}
\mathcal{L}_{\mathrm{Y}}=-\frac{\mtau}{v}\PH(\kappat\PAGt\PGt+\kappattilde\PAGt i\gamma_{5}\PGt).
\end{equation}
\end{linenomath}
In this equation, \mtau is the mass of the \PGt lepton, \PGt denotes the Dirac spinor of \PGt lepton fields, and the vacuum expectation value of the Higgs field, $v$, has a value of 246\GeV. The effective mixing angle 
\phitt for the \Yuktau coupling is defined in terms of the coupling strengths as 
\begin{linenomath}
\begin{equation}
\tan(\phitt)=\frac{\kappattilde}{\kappat},
\label{eq:phicpinkappa}
\end{equation}
\end{linenomath}
while the fractional contribution of the \CP-odd coupling \fCPtt is obtained from the mixing angle as $\fCPtt=\sin^{2}(\phitt)$. A mixing angle of $\phitt=0\,(90)^{\circ}$ corresponds to 
a pure scalar (pseudoscalar) coupling. For any other value of \phitt, the \PH has a mixed coupling 
with \CP-even and \CP-odd components, with maximal mixing  at a value of ${\pm}45^{\circ}$.

\begin{figure}[hbt]
\centering
\includegraphics[width=0.5\textwidth]{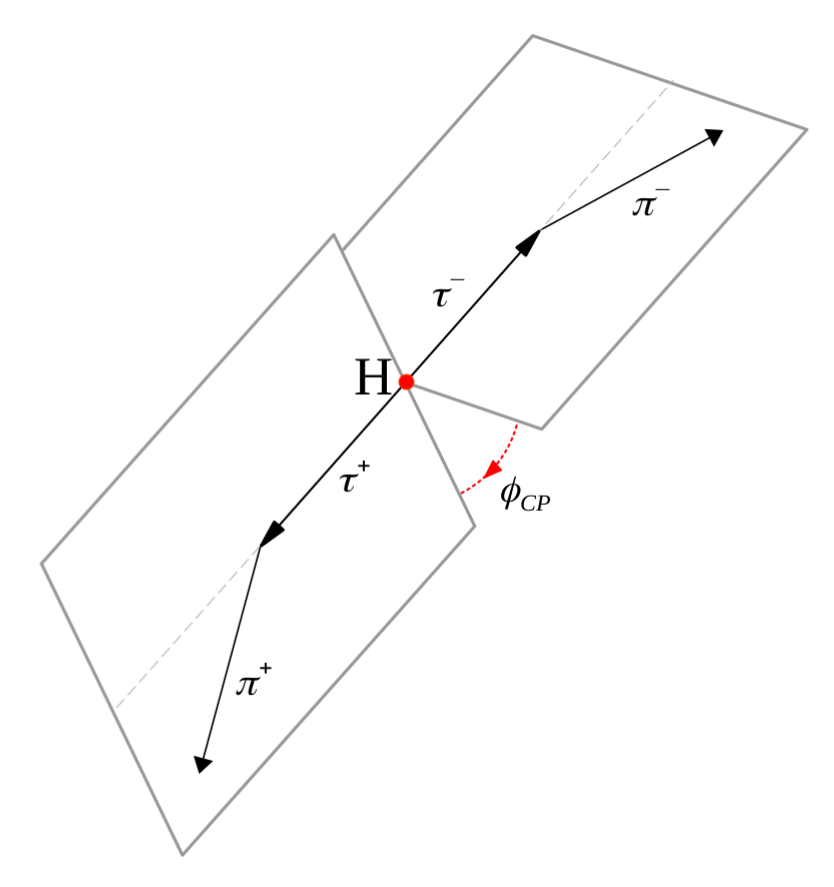}
\caption{The decay planes of two \PGt leptons decaying to a single charged pion. The angle \phicp is the angle between the decay planes. The illustration is in the \PH rest frame.
\label{fig:illustrationIP}}
\end{figure}

The angle \phicp denotes the angle between the \PGt lepton decay planes in the \PH rest frame.
An illustration of the decay planes in the single pion channel is depicted in Fig.~\ref{fig:illustrationIP}. 
The relation between \phitt and \phicp may be inferred from the decay of a \PH via \PGt leptons to two outgoing charged particles~\cite{BERGE2016841} as
\begin{linenomath}
\begin{equation}
\dd{\Gamma}{\phicp}(\PH\to \PGtp\PGtm)\sim 1-b(E^{+})b(E^{-})\frac{\pi^2}{16}\cos(\phicp-2\phitt).
\label{eq:DifTauDecayWidth}
\end{equation}
\end{linenomath}
In this equation, the outgoing charged particles have an energy $E^{\pm}$ in their respective \PGt rest frames. The functions $b$ are spectral functions~\cite{Berge:2011ij} that encapsulate the correlation between the \PGt spin and the momentum of the outgoing charged particle. 
We note that the spectral functions for the leptonic and various hadronic decays are different. 

\begin{figure}[hbt]
\centering
\includegraphics[width=0.5\textwidth]{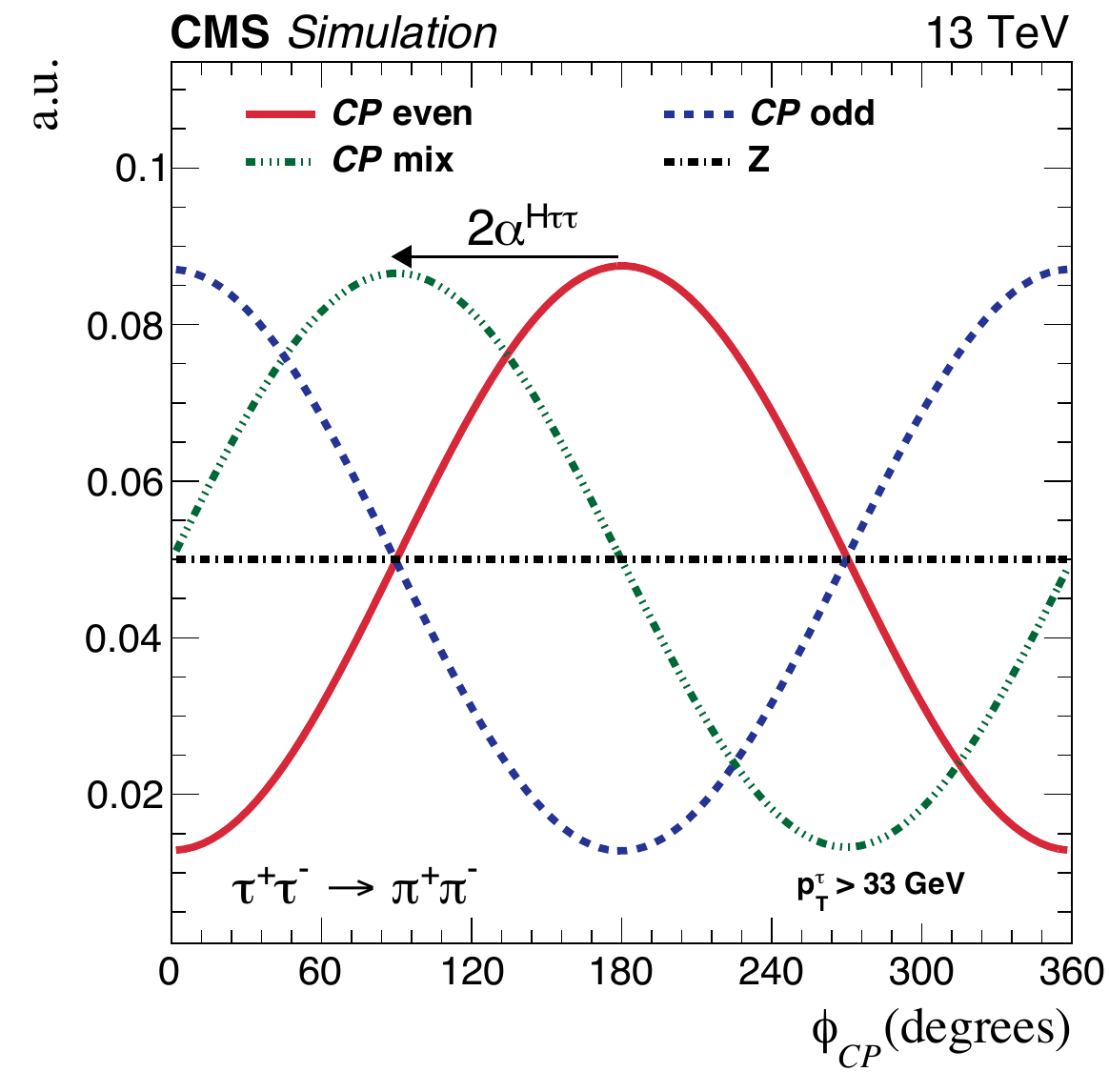} 
\caption{The normalised distribution of \phicp between the \PGt lepton decay planes in the \PH rest frame at the generator level, for both \PGt leptons decaying to a charged pion and a neutrino. The distributions are for a decaying scalar (\CP-even, solid red), pseudoscalar (\CP-odd, dash blue), a maximal mixing angle of $45^{\circ}$ (\CP-mix, dash-dot-dot green), and a \PZ vector boson (black dash-dot). The transverse momentum of the visible \PGt decay products $\pt^{\PGt}$ was required to be larger than 33\GeV during the event generation.
}
\label{fig:Introduction_Plots_CPPhases}
\end{figure}

Figure~\ref{fig:Introduction_Plots_CPPhases} shows the normalised distribution of \phicp at the generator level, calculated in the rest frame of the \PH, for the scalar, pseudoscalar, and maximally mixed values of \phitt, as well as the \phicp distribution from Drell--Yan processes. The simulated event samples that are used to generate these distributions are discussed in Section~\ref{sec:datasimulatedsamples}. These distributions are for the scenario where both \PGt leptons decay to a charged pion and a neutrino.

There is a phase shift between different mixing scenarios such that the difference in \phicp equals two times the difference in \phitt, as given by Eq.~(\ref{eq:DifTauDecayWidth}). 
It is important to note that the distribution of \phicp for the Drell--Yan background is constant; we will exploit this symmetry to reduce statistical fluctuations in the background estimates, as explained in Section~\ref{sec:eventcategorisation}.

The observable \phicp was originally introduced in the context of \EE collisions~\cite{PhysRevD.33.93,Kr_mer_1994} where the \PGt lepton momenta can be reconstructed and thus \phicp can be calculated in the \PH rest frame.  
In hadronic collisions the momenta of the neutrinos cannot be well constrained, except for the configuration in which both \PGt leptons decay via the \PaoneTHREEP mode to three charged pions---where the momenta of the \Pgt leptons can be further constrained from the reconstruction of the \Pgt lepton production and decay vertices.
Therefore, the methods for estimating \phicp have been extended and optimised for hadronic collisions~\cite{Berge2014}, as discussed in Section~\ref{sec:decayplanereco}. 
Throughout this document, we will denote the angle between the \PGt decay planes as \phicp, irrespective of the frame in which it is calculated.

\section{The CMS detector}\label{sec:CMSdetector}

The central feature of the CMS apparatus is a superconducting solenoid of 6\unit{m} internal diameter, providing a magnetic field of 3.8\unit{T}. 
Within the solenoid volume are a silicon pixel and strip tracker, a lead tungstate crystal electromagnetic calorimeter (ECAL), and a brass and scintillator hadron calorimeter (HCAL), 
each composed of a barrel and two endcap sections. Forward calorimeters extend the pseudorapidity ($\eta$) coverage provided by the barrel and endcap detectors. 
Muons are detected in gas-ionisation chambers embedded in the steel flux-return yoke outside the solenoid. Events of interest are selected using a two-tiered trigger system. The first level (L1), composed of custom hardware processors, uses information from the calorimeters and muon detectors to select events at a rate of around 100\unit{kHz} within a fixed latency of about 4\mus~\cite{Sirunyan:2020zal}. The second level, known as the high-level trigger, consists of a farm of processors running a version of the full event reconstruction software optimised for fast processing, and reduces the event rate to around 1\unit{kHz} before data storage~\cite{Khachatryan:2016bia}.
A more detailed description of the CMS detector, together with a definition of the coordinate system used and the relevant kinematic variables, can be found in Ref.~\cite{Chatrchyan:2008zzk}.

\section{Simulated samples}\label{sec:datasimulatedsamples}

The signal and relevant background processes are modelled with samples of 
Monte Carlo simulated events. The signal samples with a \PH produced through gluon-gluon fusion (\ggH), vector boson fusion (VBF),
or in association with a \PW or \PZ vector boson (denoted as $\PW\PH$ or $\PZ\PH$, or \VH when combined) are generated 
at next-to-leading order (NLO) in perturbative quantum chromodynamics (QCD) with 
the \POWHEG 2.0~\cite{Nason:2004rx,Frixione:2007vw, Alioli:2010xd, Bagnaschi:2011tu, Nason:2009ai, Jezo:2015aia, Granata:2017iod} event generator. The \PH production mechanism is configured to only produce scalar Higgs bosons, as opposed to pseudoscalar Higgs bosons or mixed couplings. The latter scenarios would also affect various properties of the production, \eg the production rate and the topology of associated jets, such as the azimuthal angle \Deltaphijj between the two leading jets, when present~\cite{Klamke:2007cu}. We note that in our analysis of \phitt we are not sensitive to the modifications to the \ggH and VBF+\VH production rates as we treat them as unconstrained parameters that are allowed to float freely in the fit to data. We also do not use the \Deltaphijj or similar variables to define event selection criteria or as inputs to discriminants; whereas modifications to other kinematic variables must be negligibly small in order to avoid experimental bounds from dedicated measurements (\eg Ref.~\cite{CMS:2021nnc}). 
For the \ggH production process, we used dedicated simulations~\cite{Demartin:2014fia,Bagnaschi:2011tu} to confirm that modifications to the \CP properties of the Yukawa couplings between the Higgs and top and bottom quarks did not significantly influence either the signal acceptance or the distributions of discriminants used to extract our results. We observed that such effects are typically at the $\mathcal{O}(1\%)$ level or smaller and are negligible compared to theoretical uncertainties on the signal modelling.
Therefore, our measurement of \phitt is not sensitive to the assumptions made about the \CP-nature of the production interactions. 

The distributions of the Higgs boson's transverse momentum (\pt) and of the jet multiplicity are reweighted to match the predictions at next-to-NLO (NNLO) accuracy obtained from full phase space calculations with the \POWHEG \textsc{nnlops} (version 1) generator~\cite{Hamilton:2013fea,Hamilton:2015nsa}.
The decay of the \PH does not depend on its production. The description of the decay of the \PH to \PGt leptons is obtained using the 
\PYTHIA generator version 8.230~\cite{Sjostrand:2014zea}. These samples are simulated without accounting for the \PGt spin correlations. 
After the samples have been generated, the \TAUSPINNER package~\cite{Przedzinski:2018ett} is used to calculate event weights that can be applied to the simulated 
signal samples to model \PGt polarisation effects for a boson with \CP-mixing angles of 0, 45, and $90^{\circ}$.
 There is no normalisation effect from the reweighting procedure, \ie the integrated \HTT cross section of the signal samples is invariant under rotations in \phitt. 
 All 2016 samples are generated with the NNPDF3.0~\cite{Ball_2015} NLO parton distribution functions (PDFs), while the NNPDF3.1~\cite{Ball_2017} NNLO distributions are used for 2017--2018. 
 
The \MGvATNLO~\cite{Alwall:2014hca} generator (version 2.6.0) is used for processes involving a \PZ or \PW boson and up to four outgoing partons generated with the matrix element, and these processes are denoted \Zjets and \Wjets, respectively. Processes involving \PW bosons originating from top quark decays are not considered in these samples.
They are simulated at leading order (LO) with the MLM jet matching and merging approach~\cite{Alwall:2007fs}.
The same generator is used at NLO for diboson production, whereas \POWHEG 
2.0 (1.0) is used for top quark-antiquark pair production~\cite{Alioli:2011as} and single top quark production (associated with a \PW boson)~\cite{Re:2010bp, Frederix:2012dh}. 
The generators are interfaced with \PYTHIA to model 
the parton showering and fragmentation, as well as the decay of the \PGt leptons. 
The \PYTHIA parameters that affect the description of the underlying event are set to the {CUETP8M1} tune~\cite{Khachatryan:2015pea} in 
2016, and {CP5} tune~\cite{Sirunyan_2020_CP5} in 2017--2018.

Monte Carlo generated events are processed through a simulation of the CMS detector
that is based on \GEANTfour~\cite{Agostinelli:2002hh}, and are reconstructed with
the same algorithms as the ones used for data.
Additional \PP interactions per bunch crossing ("pileup") are included.
The effect of pileup is taken into account by generating concurrent minimum bias 
collision events with \PYTHIA. The pileup distribution in simulation is weighted to match the pileup in data.

\section{Event reconstruction}\label{sec:eventreconstruction}

The reconstruction algorithms for both observed and simulated events are based 
on the particle-flow (PF) algorithm~\cite{Sirunyan:2017ulk}, which relies on 
the information from the different CMS subdetectors to reconstruct muons, electrons, photons,
and charged and neutral hadrons. These objects are combined to form more complex 
ones, such as \tauh candidates or missing transverse momentum (\ptmiss).

\subsection{Primary vertex reconstruction}
The positions of all \PP interactions (vertices) in the event, including the hard scatter (primary) and soft (pileup) vertices, are reconstructed in a two-step procedure~\cite{TRK-11-001}. The steps consist in clustering the tracks that appear to originate from the same interaction using the deterministic annealing algorithm~\cite{726788}, and subsequently fitting the position of each vertex using tracks associated to its cluster with the adaptive vertex fitter (AVF) algorithm~\cite{Waltenberger_2007}. The candidate vertex with the largest value of the sum of the $\pt^{2}$ of all associated physics objects is considered to be the primary \PP interaction vertex (PV). 
The physics objects included in this sum are jets, clustered using the anti-\kt jet finding algorithm~\cite{Cacciari:2008gp} with the tracks assigned to candidate vertices as inputs, and the associated \ptmiss, taken as the negative vector sum of the \pt of those jets.

\subsection{Muon reconstruction}
Muons are identified and reconstructed with requirements on the
quality of the track reconstruction and on the number of hits in the tracker and muon systems~\cite{Chatrchyan:2012xi}, and selected within $\abs{\eta}<2.4$. 
In order to reject muons that originate from nonprompt interactions, or are misidentified, a relative
isolation is defined as
\begin{linenomath}
\begin{equation}
\mIso \equiv \frac{\sum_\text{charged}  \pt + \max\left( 0, \sum_\text{neutral}  \pt
                                         - \frac{1}{2} \sum_\text{charged, PU} \pt  \right )}{\ptmu}.
\label{eq:reconstruction_isolationmu}
\end{equation}
\end{linenomath}
In this equation, $\sum_\text{charged} \pt$ is the scalar \pt sum of the charged particles originating from
the PV and located in a cone of size
$\DR = \sqrt{\smash[b]{(\Deltaeta)^2 + (\Deltaphi)^2}} = 0.4$ (where $\phi$ is azimuthal angle in radians)
centred on the muon direction. The sum $\sum_\text{neutral}  \pt$ is
a similar quantity for neutral particles. The $\sum_\text{charged, PU} \pt$ term sums over charged particles originating from pileup vertices in order to estimate and subtract the contribution of pileup to the neutral particle sum, which is scaled by $1/2$ to account for the fraction of neutral to charged energy in pileup interactions.
The \pt of the muon is denoted by \ptmu. In the \mutau channel, it is required that $\mIso<0.15$.

\subsection{Electron reconstruction}
Electrons are reconstructed using tracks from the tracking system and calorimeter deposits in the ECAL, with a veto on objects with a large HCAL to 
ECAL energy ratio. Electrons are identified using a multivariate analysis (MVA) discriminant
combining several quantities that describe the shape of the energy deposits
in the ECAL, the quality of tracks, and the compatibility of the measurements from
the tracker and the ECAL~\cite{Khachatryan:2015hwa}. 
The energy scale of electrons is adjusted in data and simulation using the \PZ mass peak, while its resolution in simulation is adjusted to data.

For the electrons, an isolation criterion \eIso is defined for a cone size of $R<0.3$ centred on the electron direction. Its definition is analogous to Eq.~(\ref{eq:reconstruction_isolationmu}) for the charged tracks, but the pileup contribution of neutral particles is estimated via an effective-area method as
\begin{linenomath}
\begin{equation}
\label{eq:reconstruction_isolationele} 
\eIso = \frac{\sum_\text{charged}  \pt  + \max\left(0, \sum_\text{neutral} \pt - \rho\,\text{EA}\right )}{\pte}.
\end{equation}
\end{linenomath}
In this equation, the pileup contribution is estimated as $\rho\,\text{EA}$, where $\rho$ is the event-specific average pileup
energy density per unit area in the $\phi$-$\eta$ plane and EA, which depends on the electron $\eta$, is the effective area specific to the neutral component of the isolation variable~\cite{Khachatryan:2015hwa}.
In the \etau channel, it is required that $\eIso<0.15$.

\subsection{Jet and \texorpdfstring{\ptmiss}{Missing transverse energy} reconstruction}
Jets are reconstructed using the anti-\kt algorithm~\cite{Cacciari:2008gp} with distance parameter $R=0.4$ as implemented in the \FASTJET package~\cite{Cacciari:2011ma}. 
The anti-\kt algorithm functions by taking PF objects and grouping them together based on 
inverse powers of the \pt of the objects~\cite{Cacciari:2008gp,CMS-PAS-JME-16-003}.
Jet momentum is determined as the vectorial sum of all particle momenta in the jet, and is found from simulation to be, on average, within 5 to 10\% of the true momentum over the whole \pt spectrum and detector acceptance. Pileup interactions can contribute additional tracks and calorimetric energy depositions to the jet momentum. To mitigate this effect, charged particles identified to be originating from pileup vertices are discarded and an offset correction is applied to correct for remaining contributions. Jet energy corrections are derived from simulation to bring the measured response of jets to that of particle level jets on average. In situ measurements of the momentum balance in dijet, $\text{photon} + \text{jet}$, $\PZ + \text{jet}$, and multijet events are used to account for any residual differences in the jet energy scale between data and simulation~\cite{Khachatryan:2016kdb}. The jet energy resolution amounts typically to 15--20\% at 30\GeV, 10\% at 100\GeV, and 5\% at 1\TeV~\cite{Khachatryan:2016kdb}. Additional selection criteria are applied to each jet to remove jets potentially dominated by anomalous contributions from various subdetector components or reconstruction failures.
Data collected in the ECAL endcaps were affected by large amounts of noise during the 2017
data-taking period, which led to disagreements between simulation and data. To mitigate this issue, 
jets used in the analysis of the 2017 data are discarded if they have $\pt < 50\GeV$ and $2.65 < \abs\eta < 3.10$.
Hadronic jets that contain {\PQb}-quarks ({\PQb}-jets) are tagged using a deep neural network (DNN), called \textsc{DeepCSV} algorithm~\cite{Sirunyan_2018DEEPCSV}. The medium working point used for the \textsc{DeepCSV} algorithm corresponds to a {\PQb}-jet identification efficiency of about 70\% for a misidentification rate for jets originating from light quarks and gluons of around 1\%.

The pileup per particle identification algorithm~\cite{Bertolini:2014bba} is applied to reduce the pileup dependence of the \ptvecmiss observable. 
The \ptvecmiss and its magnitude (\ptmiss) are computed from the PF candidates weighted by their probability to originate from the PV~\cite{Sirunyan:2019kia}.
The \ptvecmiss is adjusted for the effect of jet energy corrections.

\subsection{Tau lepton reconstruction}
The \tauh lepton reconstruction is performed with 
the \HPS (HPS) algorithm~\cite{Sirunyan_2018_TauIDCMS}. Starting from the constituents of reconstructed jets, the algorithm works by combining charged hadrons with the signature of neutral pions---one or more electron/photon candidates falling within a certain $\Deltaeta{\times}\Deltaphi$ region (referred to as a ``strip'').
The combination of these signatures provides the four-vector of the visible decay products of the
parent \tauh. The identification of \tauh candidates makes use of isolation discriminators to reject quark 
and gluon jets that could be misidentified as \tauh. For this analysis, a DNN called \textsc{DeepTau}~\cite{TAU-20-001}
is used on the HPS \tauh candidates to provide further discrimination. 
In order to achieve an optimal \tauh identification performance, the DNN combines information from the high-level 
reconstructed \tauh features together with the low-level information from the inner tracker, calorimeters and 
muon sub-detectors, using PF candidates reconstructed within the \tauh isolation cone.
The working point on the output discriminant is chosen 
to provide a \tauh identification efficiency of about 60\% at a jet misidentification rate of approximately~$5\times 10^{-3}$.
Two other DNNs are used to reject electrons and muons misidentified as \tauh candidates using dedicated criteria
based on the consistency between the measurements in the tracker, calorimeters, and muon detectors.

The mass of the \tautau system \mtt is calculated using a simplified matrix-element algorithm, \SVFIT~\cite{Bianchini_2014},
 which combines the \ptvecmiss and its uncertainty matrix with the four-vectors of both \PGt candidates
to calculate the parent boson's mass. The resolution of \mtt is 15--20\%\
depending on the \tautau final state and the boost of the \tautau system.

\section{Reconstruction of \texorpdfstring{\CP-sensitive}{CP-sensitive} observables}\label{sec:decayplanereco}

In this section we outline the methods used to construct \CP-sensitive observables, collectively referred to as \phicp angles.
Various techniques can be used to define \phicp depending on the decay topology of the \PGt leptons. In total, four methods are employed in the analysis: the ``impact parameter"~\cite{Berge:2008dr,Berge:2015nua}, ``neutral-pion"~\cite{Bower:2002zx,Berge:2015nua}, ``combined"~\cite{Berge:2015nua}, and ``polarimetric vector"~\cite{Cherepanov:2018yqb} methods. We provide a detailed description of these methods below.  
We then summarise for which di-\PGt final states each method is utilised, and outline the procedures used to optimise the resolving power of the \phicp observables.

\subsection{Impact parameter method}\label{sec:ImpactParameterMethod}
This  method exploits the finite lifetime of the \PGt leptons and can be applied to all events where both \PGt leptons decay to a single charged particle.
We define the impact parameter \vecjpm of a track (where $\pm$ refers to the charge of the track) as the vector between the PV and the point on the track where distance to the PV is minimal. 

For each \PGt lepton we define a plane using the impact parameter vector and the charged-particle momentum vector. 
This plane, which is constructed in the laboratory frame, only represents the genuine plane of the decay into a single charged pion and neutrino when the laboratory frame coincides with the rest frame of the \PH.
This means that this method does not reconstruct the genuine \PGt lepton decay plane, but rather a plane that is correlated with it. 
In order to approximate the rest frame of the \PH we use the charged decay products of the \PGt leptons of the \PH to define a zero-momentum frame (ZMF) into which the decay planes are boosted. 
The ZMF is used to define \phicp for all channels in this analysis, except the \aoneaone channel, where both \PGt leptons decay to three charged pions and the \PH rest frame can be reconstructed.
 
We then construct four-component vectors in the laboratory frame as $\lambda^\pm=(0,\vecjpm)$. The $\lambda^\pm$ four-vectors are boosted into the ZMF and denoted ${\lambdaZMF}^{\pm}$. We also boost the respective charged-pion four-vectors to the ZMF, denoted \qZMF. Subsequently, we take the transverse components of ${\lambdaZMF}^{\pm}$ with respect to \qZMF. We normalise the vectors to obtain unit vectors \lambdahatZMFplus and \lambdahatZMFminus.

\begin{figure}[htb!]
\centering
\includegraphics[width=0.3\textwidth]{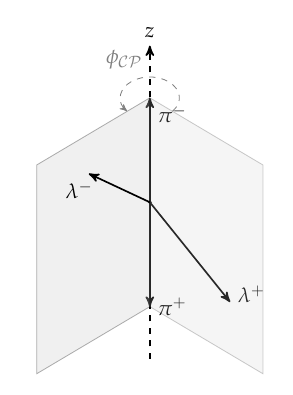}
\includegraphics[width=0.3\textwidth]{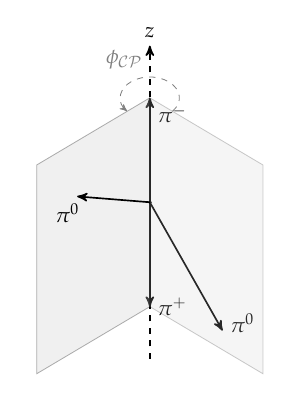}
\includegraphics[width=0.3\textwidth]{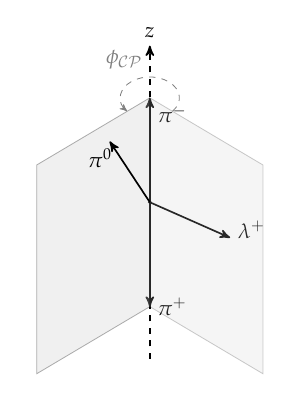}
\caption{Illustration of the \PGt lepton decay planes and the angle \phicp for various decay configurations. The decay planes are illustrated with the shaded regions, and either the vector \lambdahat or the momentum vector of the neutral pion is in the decay plane. The illustrations are in the frame in which the sum of the momenta of the charged particles is zero. Left: the decay plane for the decays $\PGtm\to\Pgpm+\PGn$ and $\PGtp\to\Pgpp+\PAGn$. Middle: the decay plane as reconstructed from the neutral and charged pion momenta. Right: \phicp for the mixed scenario, in which one \PGt lepton decays to a pion while the other decays via an intermediate \PGr meson. 
\label{fig:AnalysisStrategy_ImpactParameter}}
\end{figure}

To reconstruct \phicp, we first define the angle \phiZMF and $\OZMF$ as
\begin{linenomath} 
\begin{equation}
\begin{aligned}
\phiZMF&=\arccos(\lambdahatZMFplus\cdot\lambdahatZMFminus),\,\text{and}\\
\OZMF&=\hat{q}^{ZMF-}\cdot(\lambdahatZMFplus\times\lambdahatZMFminus).	
\end{aligned}
\end{equation}
\label{eqn:phiCP}
\end{linenomath}
From \phiZMF and \OZMF we reconstruct \phicp in a range $[0,360^{\circ}]$ as
\begin{linenomath}
\begin{equation}
\phicp =
\left\{
\begin{aligned}
		&\phiZMF  && \text{if } O^{ZMF} \geq 0 \\
		&360^{\circ}-\phiZMF && \text{if } O^{ZMF} < 0
\end{aligned}
\right..
\label{eq:phicpfromphistar}
\end{equation}
\end{linenomath}
The \PGt lepton spectral functions have opposite signs for single-pion decays and leptonic decays in the kinematic regions considered in this analysis. 
This causes a phase flip between the \phicp distributions for single pion decays and leptonic decays when the impact parameter method is used~\cite{Berge:2011ij}. 
An illustration of the definition of the \phicp observable using the impact parameter method is shown in Fig.~\ref{fig:AnalysisStrategy_ImpactParameter} (left).

\subsection{Neutral-pion method}
\label{sec:neutralpionmethod}
This method can be applied to hadronic decay channels in which both \PGt leptons undergo decays involving more than one outgoing hadron. 
We describe the method applied to the intermediate \PGr meson decay, and the intermediate \Pai meson to 1- and \mbox{3-prong} decay modes.  

For the \PGr meson decays, the vector $\lambda$ is replaced by the four-momentum vector of the \Pgpz, which means we use the planes spanned by the \PGr decay products (\eg the \Pgppm and \Pgpz in the case of the \rhotopipiz decay) to define the \phicp observable. 
The four-momentum vector of the \Pgpz is obtained as follows: to estimate the \Pgpz energy, we sum the energies of all electron/photon candidates collected by the HPS algorithm. The direction of the \Pgpz is then taken as the direction of the leading electron/photon candidate. 
In most cases the leading candidate is a photon and the direction is determined by pointing its associated ECAL clusters back towards the PV.
Finally, the mass is set to the known \Pgpz mass.

\begin{figure}[hbtp]
\centering
\includegraphics[width=0.5\textwidth]{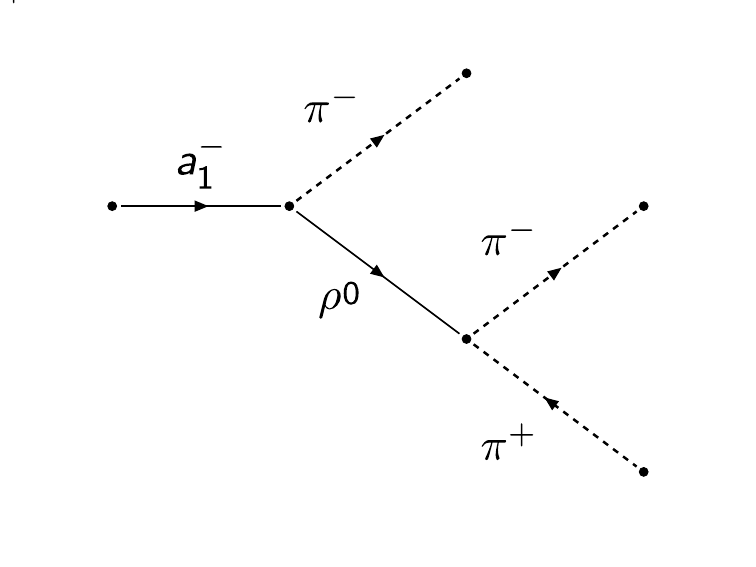}
\caption{The decay of \PaoneTHREEP via an intermediate \PGrzero to three charged pions.
\label{fig:illustration3p}}
\end{figure}

The same method is applied to \PaoneONEP decays involving two neutral pions
by summing the neutral constituents in the decay, as they cannot be easily resolved experimentally. The angle \phicp is then calculated in an analogous method to that used in the impact parameter method except that to avoid destructive interference from differently polarised states of the mesons, the following observables need to be defined:
\begin{linenomath} 
\begin{equation}
\ypm = \frac{\Epi-\Epizero }{\Epi+\Epizero }, \, \ytau = y^{\PGtm}y^{\PGtp}.
\end{equation}
\end{linenomath}
In this equation, $E_{\Pgp}$ is the energy of the pion in the laboratory frame. If \ytau is negative, \phicp is obtained via the shift $360^{\circ}-\phicp$. 
The neutral-pion method can also be successfully adapted to the \PaoneTHREEP decay mode. In these decays we select the oppositely charged pion pair with an invariant mass closest to the intermediate \PGrzero, an illustration is depicted in Fig.~\ref{fig:illustration3p}. Of this pair we treat the pion with the charge opposite of that of the \tauh lepton as though it was a \Pgpz, and the momentum of the pion with the same sign as the \tauh is used for the calculation of the ZMF.  After these assignments the neutral-pion method is applied as described for 1-prong decays.

An illustration of the definition of the \phicp observable using the neutral-pion method is shown in Fig.~\ref{fig:AnalysisStrategy_ImpactParameter} (middle).

\subsection{Combined method}

This method combines the impact parameter and neutral-pion methods outlined in the two previous sections, which is appropriate for events where only one of the two \PGt leptons decay into multiple hadrons. 
For the \PGt lepton decaying into a \PGr, \PaoneONEP, or \PaoneTHREEP mesons, the vector $\lambda$ in Eq.~(\ref{eqn:phiCP}) is replaced by four-momentum vectors as described in Section~\ref{sec:neutralpionmethod}, and the angle \phicp is then calculated using the same formulae. 

Analogously to the neutral-pion method we avoid destructive interference from differently polarised states of the mesons by applying the shift $360^{\circ}-\phicp$ for events with \ypm, where \ypm is computed for the \PGt lepton that decays to the intermediate resonance. 

An illustration of the definition of the \phicp observable using the combined method is shown in Fig.~\ref{fig:AnalysisStrategy_ImpactParameter} (right).

\subsection{Polarimetric vector method}\label{subsec:polvectors}
This method can, in principle, be applied to any \PGt lepton decay mode in which both \PGt lepton momenta can be well reconstructed. When \PGt leptons decay via the \PaoneTHREEP mode, the \PGt lepton rest frames can be reconstructed using the secondary vertices (SVs), that are extracted by fitting the three tracks originating from the \PaoneTHREEP decays.
Therefore, we only apply the polarimetric vector method to the \aoneaone decay configuration.

The polarimetric vector \vech can be considered as an estimate of the most likely direction of the spin vector $\vec{s}$ of the \PGt lepton in the \PGt lepton rest frame~\cite{Cherepanov:2018yqb}. We start by outlining the reconstruction of the \PGt lepton momenta, which are required to compute the polarimetric vectors. 
Subsequently, we describe how \phicp is reconstructed in the \PH rest frame from both the \PGt lepton momenta and the polarimetric vectors.

To reconstruct the \PGt lepton momentum in the \PaoneTHREEP channel we assume that the reconstructed \PGt lepton candidate has a mass \mtau and undergoes a two-body decay to a massless neutrino and an intermediate \Paone meson with mass \maone. Furthermore, we define the Gottfried--Jackson angle \thetaGJ as the angle between the \Paone momentum and the \PGt lepton momentum~\cite{Cherepanov:2018npf}. The latter is reconstructed from the positions of the \PGt lepton production and decay vertex. 
The magnitude of the \PGt lepton momentum is then given by~\cite{Cherepanov:2018npf}
\begin{linenomath}
\begin{equation}
\abs{\vec{p}_{\PGt}}=\frac{(\maone^2+\mtau^2)\abs{\paone}\cos\thetaGJ\pm\sqrt{\smash[b]{(\maone^2 + \abs{\paone}^2)( (\maone^2-\mtau^2)^2 -4\mtau^2\abs{\paone}^2 \sin^2 \thetaGJ )}}}{2(\maone^2+\abs{\paone}^2\sin^2 \thetaGJ) }.
\label{eq:pola_vec}
\end{equation}
\end{linenomath}
The maximal allowed value \thetaGJmax of the Gottfried--Jackson angle is defined as
\begin{linenomath}
\begin{equation}
\thetaGJmax=\arcsin\left(\frac{\mtau^{2}-\maone^{2}}{2\mtau\abs{\paone}}\right).
\label{eq:pola_vec2}
\end{equation}
\end{linenomath}
For decays in which the reconstructed \thetaGJ exceeds \thetaGJmax, due to the finite angular resolution of the charged pions and \PGt direction measurements, the value of \thetaGJ is set to \thetaGJmax.

As can be seen in Eq.~(\ref{eq:pola_vec}), there can be two solutions for the \PGt lepton momentum. This can be understood by considering the decay in the \PGt lepton rest frame. In this frame, the \Paone meson may be emitted in either the same or the opposite direction to that of the \PGt lepton momentum in the lab frame. When the \Paone meson is emitted in the direction orthogonal to the \PGt lepton, we obtain the uniquely determined solution when the square root in the numerator of Eq.~(\ref{eq:pola_vec}) vanishes. Thus, we may obtain up to four pairs of solutions for the momenta of the two \PGt leptons. 
 This ambiguity is resolved by selecting the pair of solutions with the mass closest to that of the \PH.
The direction of the \PGt lepton in the lab frame is determined by the vector SV$-$PV.

Once the \PGt leptons and \PaoneTHREEP momenta have been determined, the polarimetric vectors $\vech_{1,2}$ may be retrieved using the \PaoneTHREEP resonance model as implemented in the \TAUOLA~\cite{Jadach:1990mz,Jezabek:1991qp,Jadach:1993hs} program, which uses the parameters as measured by the CLEO Collaboration~\cite{Asner:1999kj}. 
To reconstruct \phicp from the polarimetric vectors and the \PGt lepton momenta vectors, we introduce a vector \veck that is defined as
\begin{linenomath}
\begin{equation}
\veck_{1,2}=\frac{\vech_{1,2}\times\vecn_{1,2}}{\abs{\vech_{1,2}\times\vecn_{1,2}}}.
\end{equation}
\end{linenomath}
In this definition, $\vecn_{1,2}$ are the two \PGt lepton momentum unit vectors in the \PH rest frame. 
We then reconstruct \phistar and \Ostar (in the \PH rest frame) as
\begin{linenomath}
\begin{equation}
\begin{aligned}
&\phistar=\arccos(\veck_{1}\cdot\veck_{2}), \, \text{and}\\
&\Ostar=-(\vech_1\times\vech_2)\cdot \vecn_1.
\end{aligned}
\end{equation}
\end{linenomath}
From \phistar and \Ostar we reconstruct \phicp via the assignments defined in Eq.~(\ref{eq:phicpfromphistar}).

In summary, for the configuration involving two \PaoneTHREEP decays,  the secondary decay vertices are exploited to reconstruct the \PGt momenta in the rest
 frame of the \PH.
Together with the \Paone resonance model, this allows for the extraction of the polarimetric vectors.
 Studies on simulated signal events revealed that fits to \phicp measured using the polarimetric vector method have approximately twice
 the resolving power between the \CP-even and \CP-odd states as compared to applying the neutral-pion method.
 
\subsection{Strategy for selecting the \CP-sensitive observables}
The \tauh impact parameter is relatively small compared to the tracking resolution and therefore the precision to which it can be measured is limited despite the excellent resolution of the CMS tracker.  An advantage of the neutral-pion method is that it does not rely on the reconstruction of the impact parameter; instead, the direction of the neutral pion needs to be determined.
Due to the relatively large distance between the primary interaction point and the ECAL $\left(\mathcal{O}(1\unit{m})\right)$, coupled with the fine ECAL granularity, the direction of neutral pions can be reconstructed with smaller relative uncertainties compared to the impact parameter direction.

Studies were performed on signal events to review the \CP sensitivity of the neutral pion and impact parameter methods in regions of phase space where the latter is expected to perform optimally. The sensitivity normalised to the number of events was comparable while the selections (explained below) that are needed for the impact parameter method discard a significant number of events. The cuts imposed on the impact parameter
significance mean that the neutral-pion method can be applied to about twice as many events as the impact parameter method.
Therefore, although the impact parameter method can in principle be applied to every \PGt lepton decay mode, in this analysis we only use this method for the \pipi, \mupi, and \epi final states. For the \rhorho, $\PGr\PaoneONEP$, $\PaoneONEP\PaoneONEP$, $\PaoneONEP\PaoneTHREEP$, and $\PGr\PaoneTHREEP$ final states, the neutral-pion method is deployed, and the polarimetric vector method is used exclusively for the \aoneaone channel.
In other configurations where one \PGt lepton decays to a single charged hadron or lepton and the other to multiple hadrons, the combined methods is used. 

\subsection{Extraction of \texorpdfstring{\phicp}{phicp} optimisation}
In this section we outline the experimental techniques that are developed for this analysis to optimise the experimental extraction of \phicp. A dedicated MVA discriminant is deployed to improve the identification of the \tauh decay modes.
 To improve the estimate of the impact parameters, the reconstruction 
of the PV coordinates is improved, and a helical extrapolation of the track to the PV was implemented. These methods are discussed in detail below.

\subsubsection{Multivariate discriminant for \texorpdfstring{\tauh}{tau} decay mode identification}\label{sec:decaymodemva}
In order to optimally discriminate between the different decay modes, a boosted decision tree (BDT)~\cite{CMS-DP-2020-041} is deployed. It is trained using the \textsc{XGBoost}~\cite{XGBoost} framework, and is applied on top of the \tauh selection. 
The algorithm was trained to distinguish between the 1- and 3-prong \PGt lepton decays: \Pgp, \PGr, \PaoneONEP, \PaoneTHREEP, and \THREEPPIZ. The \THREEPPIZ decay is not used in the extraction of the \CP angle but must 
be separated from \PaoneTHREEP to avoid contamination.

The inputs to the BDT are the kinematic features of the \tauh reconstructed by the HPS method and its constituents. The BDT exploits angular correlations between the decay products,
 invariant mass quantities, and kinematic properties of the photons.
 
\subsubsection{Primary vertex refitting}

The finite lifetime of the \PGt lepton means that tracks emanating from its decay do not originate from the PV. 
These tracks are removed and the PV is refitted using the remaining tracks as input to the AVF algorithm. 
The LHC beamspot represents a three-dimensional (3-D) profile of the luminous region, where the LHC beams collide in the CMS detector. The parameters of the beamspot are determined from an average over many events~\cite{TRK-11-001}. The uncertainties in the beamspot parameters are relatively small and are incorporated into the AVF algorithm to provide an additional constraint on the PV position.
The inclusion of the beamspot information leads to an improvement of the PV resolution in the transverse plane of a factor of about 3 for signal events and of about 4 for Drell--Yan events, while the $z$ coordinate of the PV is largely unaffected. 
This refitted PV is used when estimating the impact parameters and the polarimetric vectors.
 
 \subsubsection{Impact parameter estimate and significance}
 A dedicated algorithm is deployed to derive the impact parameter of the charged track from the \PGt lepton decay using an analytic extrapolation of track trajectory towards the PV position.
The extrapolation depends on the magnetic field and the helical parameters of the track.
The distance between the extrapolated track and the PV position is then minimised numerically to determine the impact parameter. 

This procedure has two advantages.   
Firstly, with this extrapolation, the minimisation of the impact parameter is performed in three dimensions. For tracks with large $\eta$ values, the procedure leads to a better estimation of the $z$ coordinate of the impact parameter than when the minimisation is done exclusively in the transverse plane.
Secondly, the helical extrapolation allows for the propagation of both the track and PV uncertainties into an overall impact parameter significance \SIP (defined as the ratio of the magnitude of the impact parameter divided by its uncertainty). In this analysis, selections are made on the impact parameter significance, as further explained in Section~\ref{sec:eventcategorisation}.

\section{Event selection}\label{sec:eventselection}

Events are selected online by the CMS trigger system. For the \ltau channels, events are triggered by either a paired \Ltau cross trigger or a single-lepton trigger with a higher \pt threshold for the lepton compared to the cross trigger.
For the \tauhtauh channel, a di-\PGt trigger is used. 

Offline, a pair of oppositely charged  \PGt leptons separated by $\DR>0.5$ is required. The offline-reconstructed objects must match the required trigger objects (\ie the object as reconstructed by the trigger system) within $\DR<0.5$. The offline-reconstructed light lepton is required to have a \pt value that is at least 1\GeV higher than the online threshold. If an offline \tauh candidate is matched to a \tauh trigger object (including the \tauh leg of the \Ltau cross  trigger for the semileptonic channels), the \tauh must have a \pt at least 5\GeV above the trigger threshold. The offline thresholds are higher than the online thresholds due to the turn-on curve in the trigger efficiencies.

Table~\ref{tab:triggertable} summarises the online trigger and offline \pt thresholds for 2016--2018. The offline requirements apply only to objects that are matched to a trigger object. 
If, in the \ltau channels, the event is selected online by the single-lepton trigger, the offline \tauh is required instead to have a \pt of at least 20\GeV.

\begin{table}[htbp]
\renewcommand{\arraystretch}{1.5}
\centering
\topcaption{Kinematic trigger and offline requirements applied to the \etau, \mutau, and \tauhtauh channels. The trigger \pt requirement is indicated in parentheses (in \GeV). The \pt thresholds indicated for the \tauh apply only for the object matched to the hadronic trigger or to the hadronic leg from the cross trigger.
\label{tab:triggertable}
}
\begin{tabular}{llll}
\hline

  Channel   &  Year &  Trigger requirement                      &Offline \pt (\GeVns{})                       \\ 
  \hline 
\tauhtauh   & All years & $\tauh (35)\,\&\,\tauh (35)$       & $\pttauh>40$    \\
 \multirow{2}{*}{\mutau}  & 2016	   & $\Pgm(22)$, $\Pgm(19)\,\&\,\tauh (20)$             & $\ptmu>20$, $\pttauh>25$        \\
              & 2017, 2018 & $\Pgm(24)$, $\Pgm(20)\,\&\,\tauh (27)$              &   $\ptmu>21$, $\pttauh>32$     \\
 \multirow{3}{*}{\etau}  & 2016	   & $\Pe(25)$  			              & $\pte>26$         \\
              & 2017           & $\Pe(27)$, $\Pe(24)\,\&\,\tauh (30)$           &   $\pte>25$, $\pttauh>35$      \\
              & 2018           & $\Pe(32)$, $\Pe(24)\,\&\,\tauh (30)$           &   $\pte>25$, $\pttauh>35$      \\
\hline
\end{tabular}
\end{table}

For the \ltau channels, the large \Wjets background is reduced by rejecting events based on the transverse mass \mT of the light lepton and \ptvecmiss system,
\begin{linenomath}
\begin{equation}
\mT \equiv \sqrt{{2 \pt^{\ell} \ptmiss [1-\cos(\Deltaphi)]}} < 50\GeV,
\label{eq:transversemass}
\end{equation}
\end{linenomath}
where \Deltaphi is the azimuthal angle between the direction of the light lepton and \ptvecmiss.

The longitudinal and transverse impact parameters \dz and \dxy of the muon and electron are required to satisfy $\abs{\dz}<0.2\cm$ and $\abs{\dxy}<0.045\cm$. These impact parameters originate from a minimisation of the magnitude of the 
impact parameters in the transverse plane only, in contrast to the impact parameters used for calculating \phicp, which are derived using a 3-D minimisation. For the purpose of the event selection this factorised approach is sufficiently precise. For the leading 
\tauh track, only the requirement $\abs{\dz}<0.2\cm$ is imposed to avoid loss of selection efficiency. Further, a veto on events containing loosely identified additional electrons or muons is imposed. For the \ltau channels, a veto on 
jets passing {\PQb}-tagging requirements is also applied. When multiple \PGt lepton pairs are present, the pairs are ranked based on the output scores of the \textsc{DeepTau} algorithm for the \tauh candidates, and the relative isolation for the \taul candidates. The highest ranked pair is selected. 

\section{Background estimation}\label{sec:backgroundestimation}

The processes that contribute to the background in this analysis are \Zjets, \Wjets, top quark-antiquark pair production (\ttbar), single top quark, and diboson production.
Additionally, events comprised uniquely of jets produced through the strong interaction, referred to as QCD multijet events, form a significant background. 
These processes contribute to the production of genuine \PGt leptons, jets and leptons that are misidentified as \tauh, as well as prompt leptons and jets that are misidentified as \taul in the semileptonic channels.
All background processes resulting in two genuine \PGt leptons constitute a major background, and are estimated from data using a \PGt-embedding technique~\cite{Sirunyan_2019_Embedding}. 
The majority of the backgrounds due to jets misidentified as \tauh candidates are estimated using the ``fake factor" (\FF) method, as described in Ref.~\cite{Sirunyan:2018qio}.
The remaining minor backgrounds are determined using simulated events. In the remainder of this section we outline the \PGt-embedding and \FF methods, and describe the corrections applied to simulated events in order to improve their description of the data.

\subsection{The \texorpdfstring{\PGt}{Tau} embedding method}
\label{sec:embedding}
In order to obtain the genuine \tautau background we exploit lepton universality, and replace oppositely charged muon pairs in data events with simulated oppositely charged \PGt lepton pairs. The dominant process for this background is \ZTT, but there are also small contributions from \ttbar and diboson processes.

For all data-taking periods, events containing an oppositely charged dimuon pair were collected using a dedicated di-\Pgm trigger. The detector hits belonging to the muon tracks are removed from these events. A \PZ boson is simulated in an empty detector, which is forced to decay to a pair of oppositely charged \PGt leptons with identical kinematics to the muon pair that was removed. The \PGt leptons are forced to decay fully hadronically or semileptonically in order to simulate either the \tauhtauh or \ltau channels. The detector response to the \PGt pair is then simulated and added to the data event.

In order to model the background processes in data well, various corrections need to be applied to the embedded event samples.
Muons and electrons are corrected for mismodelling of their trigger, tracking and identification, and isolation requirements efficiencies. The \tauh candidates are corrected for mismodelling of their trigger, reconstruction and identification efficiencies. The ``tag-and-probe" method~\cite{Khachatryan_2011} is used to derive these corrections. 
The \tauh energy scales are corrected per decay mode to match the corresponding scales in data.
The electron, muon, and pion impact parameters are corrected using a samples of \Zmm and \Zee events and quantile-mapping techniques. 

\subsection{The \texorpdfstring{\FF}{fake factor} method}
This  method is designed to provide an estimate of the shape and normalisation of events in which at least one quark or gluon jet is misidentified as a \tauh lepton based on control samples in data. We refer to such a jet as a \jettotauh. 

We define a determination region that is orthogonal to the signal region and dominated by a background process resulting in \jettotauh misidentifications; the construction of these regions is outlined below.
We define a \tauh nominal ID as a \tauh object that passes nominal ID requirements as outlined in Section~\ref{sec:eventreconstruction}, and a relaxed \tauh ID as objects that pass a looser requirement on the DNN output but fail the nominal ID.
In this determination region we obtain the ratio between the nominal ID \tauh rate and the relaxed ID \tauh rate. The ratio in the determination region is the \FF.
To obtain the rate of misidentified jets in the signal region, an application region is defined by selecting events that fulfil all event selection criteria except that they
 contain a \tauh lepton that passes the relaxed instead of the nominal requirement (for the \tauhtauh channel it must be the leading \tauh). The rate of misidentified jet events in the signal region is obtained by applying the \FF values from the determination region on an event-by-event basis as an event weight to the events in the application region.
In both determination and application regions the contribution of other background processes not involving \jettotauh events, which amounts to about 1\% (5\%) in the \tauhtauh (\ltau) channel(s), is subtracted using simulated events. The contamination from signal events is significantly smaller ($<$0.1\%) and is therefore neglected.

The \jettotauh background in the \tauhtauh channel originates almost entirely from QCD multijet events. The determination region is thus defined by inverting the opposite-sign requirement on the \PGt lepton pair to a same-sign requirement, which effectively selects a control region pure in QCD jets. The \FF are parameterised for the leading \tauh lepton as a function of the \pt of the \tauh, and binned in the reconstructed decay mode, jet multiplicity, and impact parameter significance. 
Correction factors are derived using control regions in data to correct for residual differences in the \ptvecmiss spectrum, and to account for the sign inversion used to define the determination region.
The final \FF value for the \tauh channel is obtained by applying the raw \FF and the two corrections multiplicatively. This \FF also accounts for other processes with a jet misidentified as the leading \tauh lepton, such as \Wjets production. The events in which the subleading \tauh is a misidentified jet and the leading \tauh candidate is a genuine \PGt lepton are modelled via simulation; these events constitute only a small fraction $(\mathcal{O}(2\%))$ of the total misidentified jet background in the \tauhtauh channel.

In the \ltau channels, the \Wjets process, and to a lesser extent the \ttbar process, contribute to jet misidentification as well as events originating from QCD multijet production.
Therefore, separate \FF are derived for these processes, and these individual \FF values are subsequently weighted into an overall \FF to account for their different contributions in the application region.
Simulated events are used to determine the expected relative contributions of \Wjets and \ttbar events, and the QCD contribution is estimated by subtracting all simulated non-QCD processes from the data in the application region. 
In order to account for dependencies of the weights on several kinematic variables, a multi-class BDT is trained to separate \Wjets, QCD, and \ttbar events. The inputs to the BDT include kinematic features of the reconstructed \tautau system and the associated jets, as well as the \tauh decay mode. The output of the BDT is a set of three scores (one per class) that sum to unity---meaning one of the outputs is redundant.
The weights are thus determined in bins of two of these scores. 
The overall \FF accounts for the jet misidentification in all background processes. The procedure for the QCD \FF is similar to the method described for the \tauhtauh channel, except that the light lepton isolation parameter must be larger than 0.05 to reduce processes resulting in genuine leptons. Correction factors are derived to correct for residual differences in the lepton \pt and \ptvecmiss spectra, and to account for the sign inversion and additional lepton isolation requirement used to define the determination region. A determination region sufficiently pure in \Wjets is defined by selecting events with $\mT>70\GeV$. Correction factors are derived to correct for residual differences in the lepton \pt and \ptvecmiss spectra, and to account for the inverted \mT selection used to define the determination region. For the \ttbar process, it is difficult to define a sufficiently pure region in data, and thus the \FF values are estimated from a simulated \ttbar sample. 
Correction factors are derived to correct for residual differences in the \ptvecmiss spectrum, and to account for differences in the \FF in data and simulation. The latter is derived by comparing the \FF values measured for \Wjets in data and simulation. 

\subsection{Estimation of minor backgrounds}
The \PGt-embedding and \FF methods combined describe around 90\% of the backgrounds in this analysis. All events containing a genuine \PGt lepton pair are taken from the embedded samples, while events in which the (leading) \tauh is a misidentified hadronic jet in the (\tauhtauh) \ltau channels are obtained from data using the \FF method. All other background events are obtained from simulation. 

In addition to the genuine \PGt and \jettotauh contributions to the selected pairs, there are additional sources of \PGt misidentifications that may occur. This includes prompt leptons that may be either misidentified as a \taul or as a \tauh, \taul leptons being misidentified as \tauh, and jets misidentified as \taul candidates. In Tables~\ref{tab:bgcontributionstt} and \ref{tab:bgcontributionslt} we summarise the different background composition configurations and their modelling for the \tauhtauh and \ltau channels, respectively. To avoid double-counting events with a genuine \PGt lepton pair, such events are subtracted from all simulated samples, as well as events in which the \tauh is a misidentified hadronic jet (for the \tauhtauh channel this must be the leading \tauh).

\begin{table}[!ht]
\topcaption{The different sources of backgrounds in the \tauhtauh channel are shown in the rows and columns. The entries in the table represent the possible \PGt lepton pair background contribution from different processes and misidentifications and encapsulate the different experimental techniques that are deployed to estimate the background contributions.}
\label{tab:bgcontributionstt}
\centering
 \begin{tabular}[t]{lccccc}
\hline
Leading \tauh			&  \multicolumn{3}{c}{Subleading \tauh} \\
					& Genuine \tauh		   & Jet \totauh & (Prompt lepton/\taul) \totauh	   \\
 \hline
  Genuine \tauh 	 		&   \PGt-Embedding      & Simulation		     &	Simulation			\\
  Jet \totauh  	&   \FF            & \FF	     &	\FF		\\
  (Prompt lepton/\taul) \totauh	&   Simulation              & Simulation	     	     & Simulation			\\
\hline
\end{tabular}
\end{table}

\begin{table}[!ht]
\topcaption{The different sources of backgrounds in the \ltau channel are shown in the rows and columns. The entries in the table represent the possible \PGt lepton pair background contribution from different processes and misidentifications and encapsulate the different experimental techniques that are deployed to estimate the background contributions.}
\label{tab:bgcontributionslt}
\centering
 \begin{tabular}[t]{lccccc}
\hline
\taul			&  \multicolumn{3}{c}{\tauh} \\
					& Genuine \tauh		   &Jet \totauh &(Prompt lepton/\taul) \totauh 	   \\
 \hline 
 Genuine \taul 	 		&   \PGt-Embedding      & \FF	     &	Simulation			\\
 Jet $\to\taul$  	&   Simulation           	   & \FF	     & Simulation			\\
 Prompt lepton $\to\taul$	&   Simulation              & \FF	     	     & Simulation			\\
\hline
\end{tabular}
\end{table}

In order to model the background processes in data well, various corrections need to be applied to the simulated samples. 
All corrections to the \PGt lepton decay products applied to the embedded samples (described in Section~\ref{sec:embedding}) are also applied to the simulated samples. 
Although both embedded and simulated samples include simulated leptons, the corresponding corrections can differ slightly due to deposits from other nearby objects, that may influence, for example, isolation sums and/or particle identification decisions.  Therefore, dedicated correction factors are derived in each case.

Jet energy scale corrections are applied to both data and simulated events as described in Section~\ref{sec:eventreconstruction}.
Recoil corrections to the \ptvecmiss are applied to reduce the mismodelling of the simulated
\Zjets, \Wjets, and Higgs boson samples. The corrections are applied to the simulated events based on the vectorial difference
of the measured \ptmiss and total \pt of the neutrinos originating from the decay of the \PZ, \PW, or \PH. Their average effect is the reduction of the \ptmiss obtained from simulation by a few \GeV. Recoil corrections to \ptvecmiss are measured in \Zmm events. The corrections are subsequently applied to Drell--Yan plus jets events, \Wjets, and signal event samples.
The \ltotauh misidentification rates are corrected in simulation by applying the tag-and-probe method to \Zll events, and the energy scales are corrected in simulation to match the scale in data.

The \PZ boson mass and \pt spectra in simulation are corrected to better match the data. To this purpose the \PZ mass and \pt are measured in data and simulation in di-muon events, and the simulated events are corrected to match the spectra in data. A correction is also applied to the top quark \pt spectrum in the \ttbar sample, using a dedicated control region. The procedure used to derive this correction is detailed in Ref.~\cite{Khachatryan_2015}.

After applying all corrections, we obtain a satisfactory description of the observables that we use to categorise events, which are described in Section~\ref{sec:eventcategorisation}.

\subsection{Validating the modelling of the \texorpdfstring{\phicp}{phicp} observables}\label{sec:backgroundvalidation}
\label{ref:validate_model}

\begin{figure}[hbtp]
\centering
\includegraphics[width=0.49\textwidth]{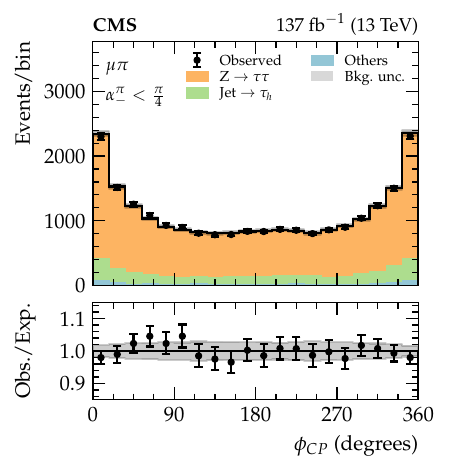} 
\includegraphics[width=0.49\textwidth]{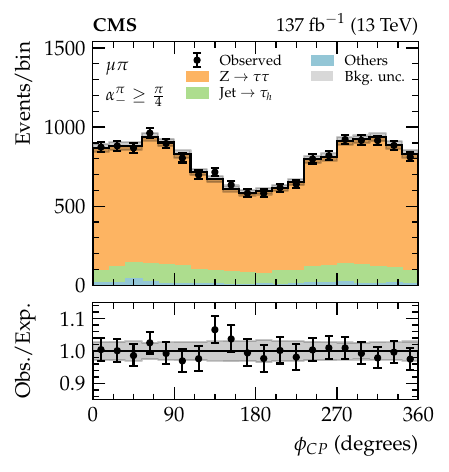} \\
\includegraphics[width=0.49\textwidth]{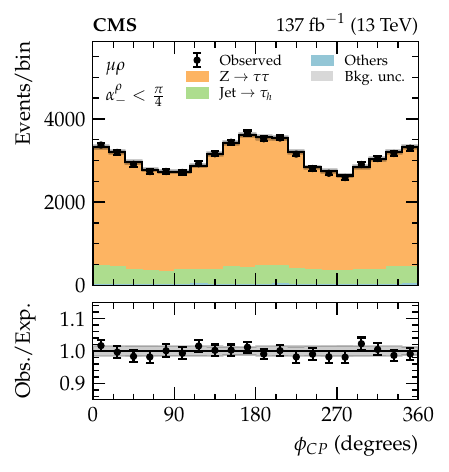}
\includegraphics[width=0.49\textwidth]{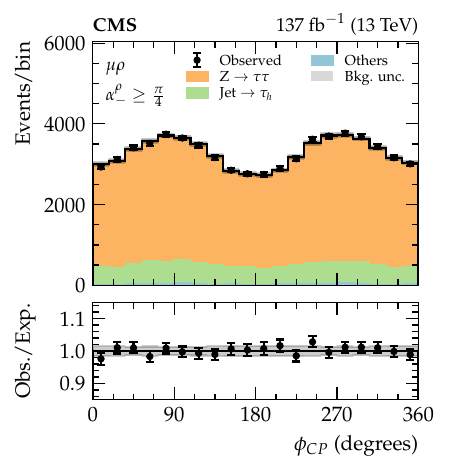} 
\caption{The angle \phicp for \mutau  events in which the \tauh decays to a charged \Pgp (upper) or a charged \PGr meson (lower). The distributions are decomposed in a subset in which the charged \Pgp is ``nearly coplanar" (left) or ``nearly perpendicular" (right) to the production plane. }
\label{fig:decomposedDY}
\end{figure}

To validate the modelling of the \phicp spectrum, the \mutau events in data and the background estimates are divided into distributions in which the charged \Pgp is ``nearly coplanar" or ``nearly perpendicular" to the production plane of the beam axis and the \PGt momentum in the laboratory frame, as described in Ref.~\cite{Berge2014}. To this purpose we introduce the observable \alphapi that is defined as
\begin{linenomath}
\begin{equation}
\cos{\alphapi}=\left|\frac{\zhat\times \phat }{\abs{ \zhat\times \phat }} \cdot \frac{\jhat\times \phat }{\abs{ \jhat\times \phat }}\right|.
\label{eqn:alpha}
\end{equation}
\end{linenomath}
In this equation, \zhat is the unit vector pointing along the beam axis, \phat is the unit momentum vector of the charged \Pgp, and \jhat is the unit impact parameter vector. We can define a subset of events in which the charged \Pgp is nearly perpendicular or coplanar by requiring $\alphapi>\pi/4$ or $\alphapi<\pi/4$, respectively.
We also perform equivalent checks for \PGt decays into \PGr mesons, where we substitute the unit \Pgpz momentum vector for \jhat in Eq.~(\ref{eqn:alpha}) to define an equivalent observable, \alpharho.
In Fig.~\ref{fig:decomposedDY} we display the coplanar and perpendicular distributions in the \mupi and \murho channels. 

\section{Event categorisation}\label{sec:eventcategorisation}

In order to enhance the sensitivity of this analysis we apply MVA discriminants to separate signal from background events.

This event categorisation is formulated as a multi-class problem.  
The output is a set of scores for each event (one per class) that, by construction, sum to unity. 
Each score can therefore be interpreted loosely as the probability that an event belongs to a given class.
We then assign each event to a category depending
on the class that received the highest score.
Since both the \ltau and \tauhtauh channels are dominated by backgrounds containing contributions
from genuine \PGt and \jettotauh production, the discriminant is trained to categorise  events in three classes:

\begin{itemize}
\item The ``Higgs" category is trained to distinguish events from the \ggH, VBF, and \VH samples from background events, which are reweighed by their cross sections before merging them into one sample. Events in this category are used to infer the \CP quantum number of the boson.
\item The ``Genuine" category includes all background processes involving two genuine \PGt leptons.
\item The ``Mis-ID" category includes all background processes in which minimally one hadronic jet is misidentified as a \tauh lepton. This category also contains \ltotauh misidentified events for all channels and prompt light leptons in the \ltau channels.
\end{itemize}

The three categories are mutually exclusive and, by definition, the lower bound for the highest MVA score is 1/3.
Subsequently, the three training categories are normalised to account for the different number of events in each data set. All event classes are then chosen to contribute to the
training with the same weight, \ie with uniform prevalence. For the semileptonic channels, the backgrounds for the training are provided from simulated samples, except for QCD events, which are obtained using same-sign \PGt lepton pair candidates in data. For the hadronic channel, the embedded samples and the \FF method are used in the training. For the latter, the events from the application region are used and reweighted by their \FF values.
The contribution of other background processes not involving \jettotauh events is not removed in this case, but the impact of these events on the performance of the MVA is negligible as they amount to only $\mathcal{O}(1\%)$ of the total.  
A separate training was performed for each year to account for differences in the performance of the CMS detector in different data-taking periods.
In the \ltau channels, the event categorisation is performed with a multiclass neural network. 
In the \tauhtauh channel, the event categorisation is performed using a multi-class BDT algorithm combined with the \textsc{XGBoost} package.
The input variables used in the categorisation of the \ltau and \tauhtauh channels are displayed in Table~\ref{tab:catvars}. The training is performed inclusively for all the \PGt lepton decay modes.  

Events are sorted into the three categories depending on which of the three output scores is closest to unity. The maximum output score is also retained and used for the purpose of signal extraction. These maximum scores will be referred to as the ``MVA scores" henceforth.   

After the categorisation, a cutoff of $\SIP>1.5$ is applied to the impact parameter significances of the electron and muon, as well as to the single pions from a \PGt lepton for events that are classified as signal events. Events 
with a lower \SIP would dilute the sensitivity of the analysis. In the background categories, a cutoff on the impact parameter significance is only applied to the single-pion decays.

\begin{table}[htb]
\topcaption{Input variables to the MVA discriminants for the \ltau and \tauhtauh channels. The \SVFIT algorithm is used to estimate the di-\PGt mass.}
\label{tab:catvars}
\centering
\begin{tabular}{lcr}
\hline
Observable & \ltau  & \tauhtauh \\
\hline
\pt of leading \tauh &$\checkmark$	&$\checkmark$   \\
\pt of trailing \tauh & \NA	&$\checkmark$   \\
\pt of \taul &$\checkmark$	& \NA  \\
\pt of visible di-\PGt & $\checkmark$		&$\checkmark$  \\
\pt of  di-\tauh + \ptmiss	& \NA	&$\checkmark$  \\
\pt of \ltau + \ptmiss & $\checkmark$	& \NA \\
Visible di-\PGt mass	& $\checkmark$		&$\checkmark$  \\
Di-\PGt mass (using \SVFIT)	& $\checkmark$ &$\checkmark$  \\
Leading jet \pt	& $\checkmark$		&$\checkmark$  \\
Trailing jet \pt	& $\checkmark$		& \\
Jet multiplicity	& $\checkmark$		&$\checkmark$  \\
Dijet invariant mass	& $\checkmark$		&$\checkmark$  \\
Dijet \pt	& $\checkmark$		& \\
Dijet $\abs{\Deltaeta}$	& $\checkmark$		& \\
\ptmiss	& $\checkmark$		&$\checkmark$  \\
\hline
\end{tabular}
\end{table}

\begin{figure}[htb]
    \centering
       \includegraphics[width=0.49\textwidth]{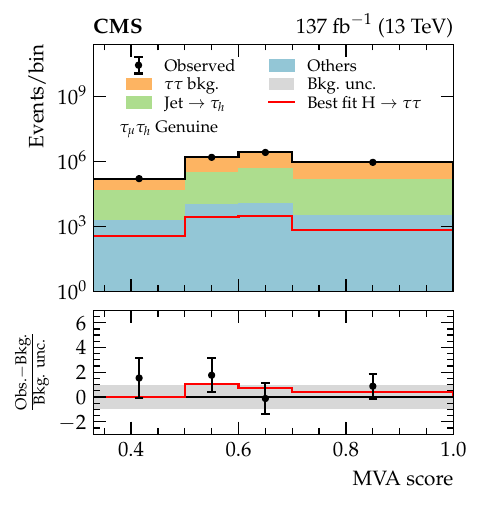}
       \includegraphics[width=0.49\textwidth]{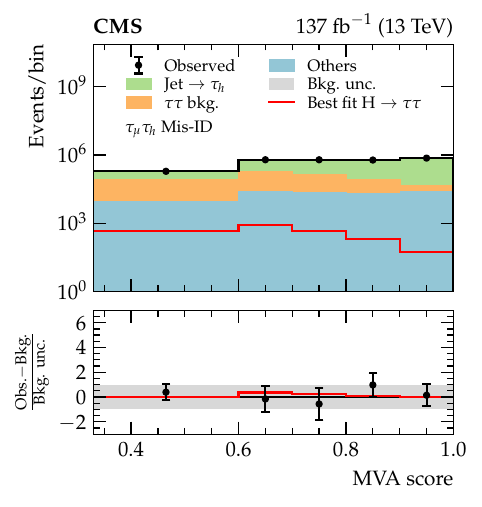}\\
       \includegraphics[width=0.49\textwidth]{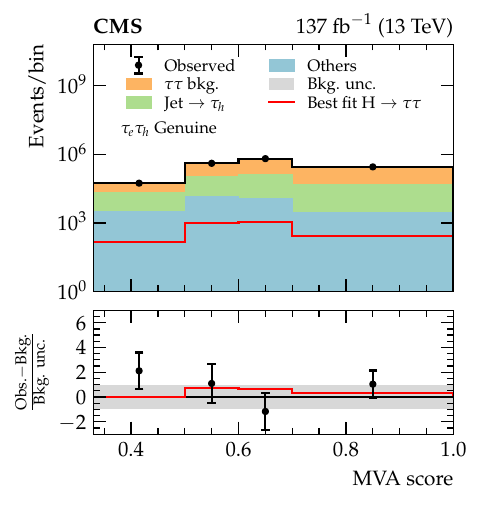}
       \includegraphics[width=0.49\textwidth]{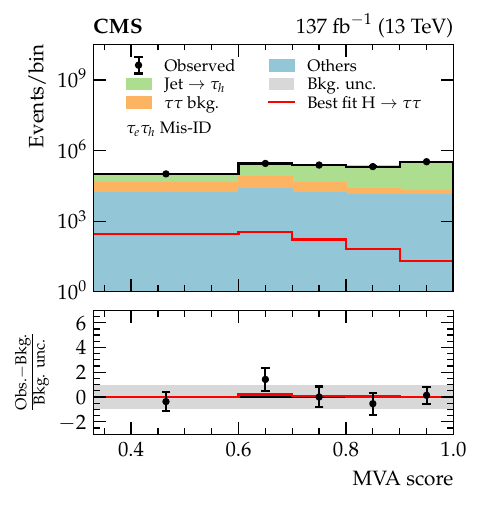} 
    \caption{The post-fit MVA score distributions for the Genuine (left) and Mis-ID categories (right) in the \mutau (upper) and \etau (lower) channels. The distributions
    are inclusive in \tauh decay mode. The best fit signal distributions are overlaid, where the signal cross sections are set to the values obtained from the fit to data, which are given in Section~\ref{sec:alpha_results}. In the lower panels, the data minus the background template divided by the uncertainty in the background template is displayed, as well as the signal distribution divided by the uncertainty in the background template.  The uncertainty band accounts for all sources of systematic
uncertainty in the background prediction, after the fit to data.\label{fig:mt_NN}}
\end{figure}

\begin{figure}[htb]
    \centering
       \includegraphics[width=0.49\textwidth]{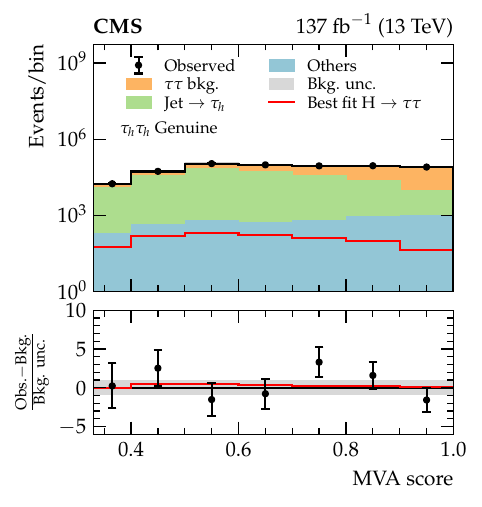}
       \includegraphics[width=0.49\textwidth]{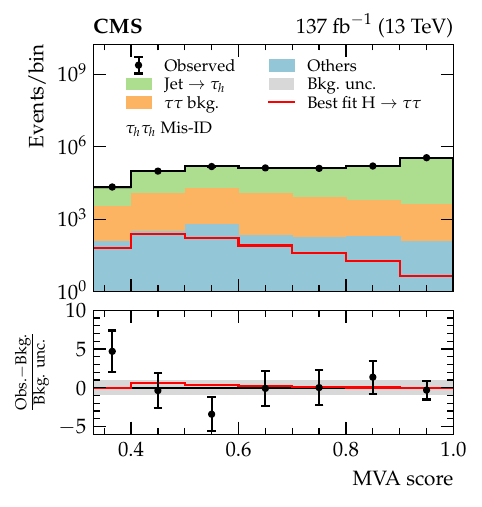}
    \caption{The post-fit MVA score distributions for the Genuine (left) and Mis-ID categories (right) in the \tauhtauh channel. The distributions
    are inclusive in \tauh decay mode. The best fit signal distributions are overlaid, where the signal cross sections are set to the values obtained from the fit to data, which are given in Section~\ref{sec:alpha_results}. In the lower panels, the data minus the background template divided by the uncertainty in the background template is displayed, as well as the signal distribution divided by the uncertainty in the background template.  The uncertainty band accounts for all sources of systematic
uncertainty in the background prediction, after the fit to data.\label{fig:tt_BDT}}
\end{figure}

In Fig.~\ref{fig:mt_NN}, the post-fit MVA score distributions of the Genuine and Mis-ID categories are displayed for the \mutau and \etau channels. The best fit signal contributions are overlaid. The fitting procedure is outlined in Section~\ref{sec:results}.
The genuine di-\PGt and \jettotauh background contributions are displayed separately as indicated in the legends. 
The remaining background contributions are collated and indicated by the ``Others" label.
The BDT scores for the \tauhtauh channel are analogously displayed in Fig.~\ref{fig:tt_BDT}.

\section{The \texorpdfstring{\phicp}{phicp} distributions in windows of the MVA discriminant score}\label{sec:unrolled}
The MVA score distributions described in Section~\ref{sec:eventcategorisation} allow for a partial separation of signal from background events. 
The \phicp distributions of the events in the signal categories are then analysed in windows of increasing MVA score, corresponding to progressively higher signal-to-background ratios.
The result is a set of 2-D distributions built from the MVA score and \phicp variables. 
These distributions are used in the fit to data to extract the results. 

The statistical fluctuations in the estimates of the background contributions (denoted as background templates) in the signal and background categories are sizeable.
It has been underlined that backgrounds involving two genuine \PGt leptons are flat in \phicp at the generator level~\cite{Berge2014}. Experimental smearing effects do not modulate this flat shape for decay modes in which we apply the neutral pion method for at least one \PGt lepton. Therefore, for this background process and these decay modes we flatten the background templates by merging the bins. The \phicp distribution is not flat for the \jettotauh background for all decay modes due to the kinematic properties of the events, but the distributions are still symmetric around $\phicp=180^{\circ}$, and so this background is symmetrised---meaning the symmetry around $\phicp=180^{\circ}$ is enforced. For other background templates, for example the \mutotauh contribution, the distributions are found to be flat within the statistical uncertainties, and therefore these background templates are also flattened. 

The backgrounds are not expected to be flat in decay modes in which the impact parameter method is used or when the polarimetric vector method is applied when both \PGt leptons decay via the \PaoneTHREEP mode. This can be understood from the fact that smearing effects in the PV  are correlated for the decay planes. The smearing of the PV results in a depletion in the region $\phicp=180^{\circ}$~\cite{Berge2014}, such that the shape of the background distributions in the \mupi, \epi, and \aoneaone (\pipi) channels tends to resemble the \CP-even (\CP-odd) signal rather than the \CP-odd (\CP-even). However, for such channels the backgrounds are symmetric around $\phicp=180^{\circ}$, and therefore the background templates are symmetrised.

For certain decay modes, the statistical fluctuations in the signal templates are also sizeable. Therefore, the templates for the scalar and pseudoscalar cases are symmetrised around $\phicp=180^{\circ}$ as well. The maximally mixed signal template, which is not displayed in the plots, is used in the fitting procedure described in Section~\ref{sec:results}. In order to symmetrise this template, we reweight the signal sample to another sample with $\phitt=-45^{\circ}$. The \phicp distribution is shifted by $180^{\circ}$, and the average is taken between the sample with $\phitt=45^{\circ}$ and $\phitt=-45^{\circ}$. 

\begin{figure}[htb]
    \centering
     \includegraphics[width=1\textwidth]{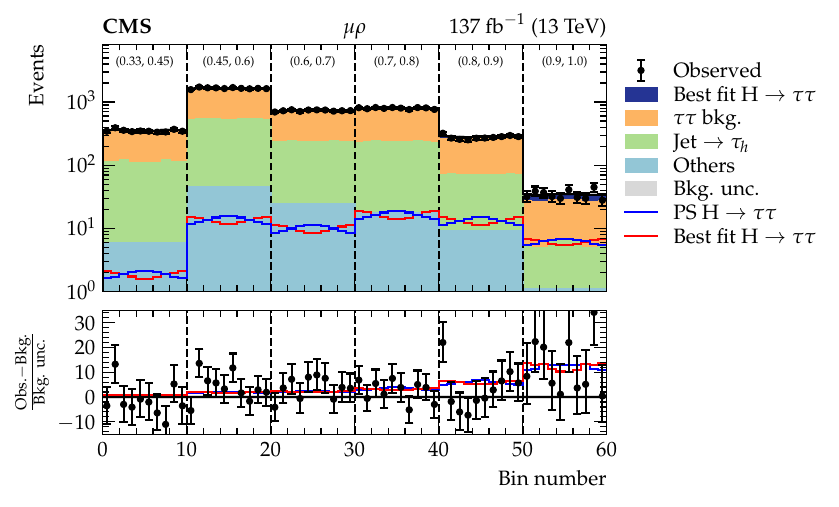}
     \includegraphics[width=1\textwidth]{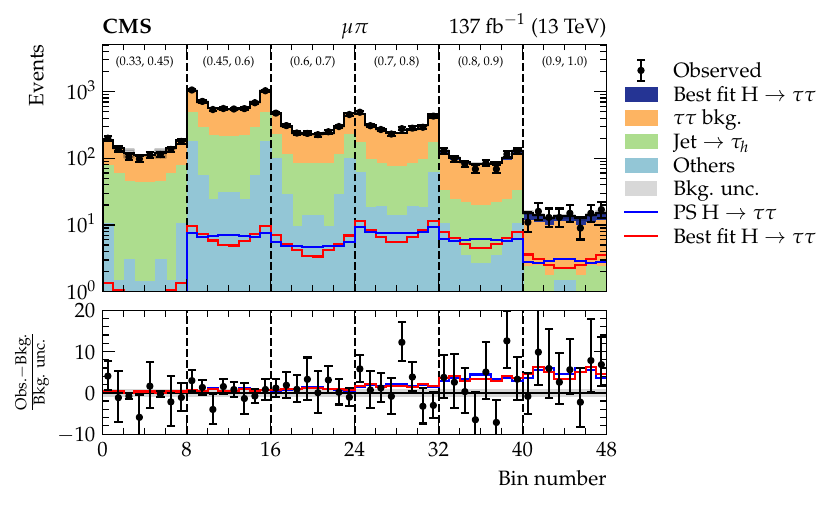}
\caption{Distributions of \phicp in the \murho (upper) and \mupi (lower) channels in windows of increasing MVA score, shown on top of each window. The best fit and pseudoscalar (PS) signal distributions are overlaid, where in both cases the signal cross sections are set to the values obtained from the fit to data, which are given in Section~\ref{sec:alpha_results}. The $x$-axis represents the cyclic bins in \phicp in the range of $(0, 360^{\circ})$. In the lower panels, the data minus the background template divided by the uncertainty in the background template is displayed, as well as the signal distributions divided by the uncertainty in the background template. The uncertainty band accounts for all sources of systematic
uncertainty in the background prediction, after the fit to data.\label{fig:unrolledmt}}
\end{figure}

\begin{figure}[htb]
    \centering
     \includegraphics[width=1\textwidth]{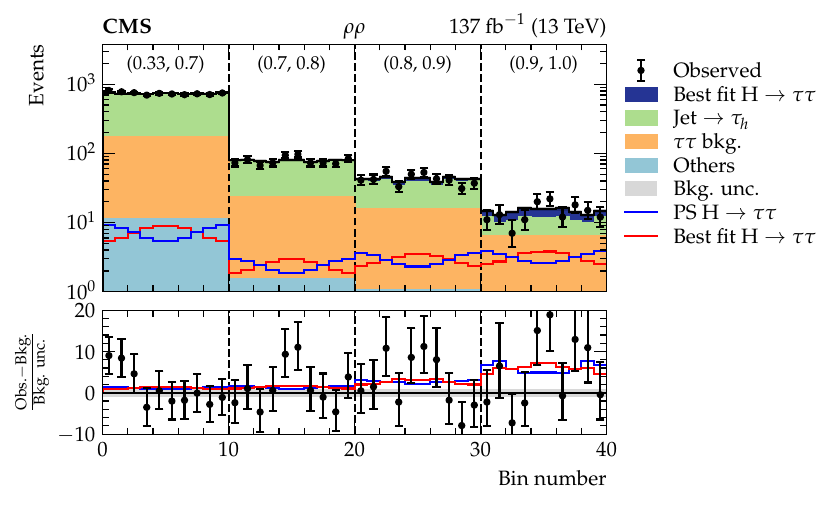}
     \includegraphics[width=1\textwidth]{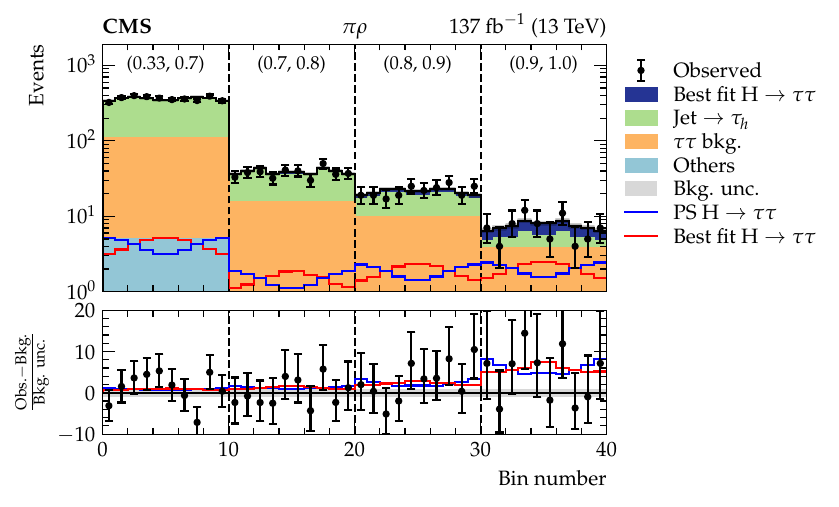}
\caption{Distributions of \phicp in the \rhorho (upper) and \pirho (lower) channels in windows of increasing MVA score, shown on top of each window. The best fit and pseudoscalar (PS) signal distributions are overlaid, where in both cases the signal cross sections are set to the values obtained from the fit to data, which are given in Section~\ref{sec:alpha_results}. The $x$-axis represents the cyclic bins in \phicp in the range of $(0, 360^{\circ})$. In the lower panels, the data minus the background template divided by the uncertainty in the background template is displayed, as well as the signal distributions divided by the uncertainty in the background template. The uncertainty band accounts for all sources of systematic
uncertainty in the background prediction, after the fit to data.\label{fig:unrolledtt}}
\end{figure}

\begin{figure}[htb]
    \centering
     \includegraphics[width=1\textwidth]{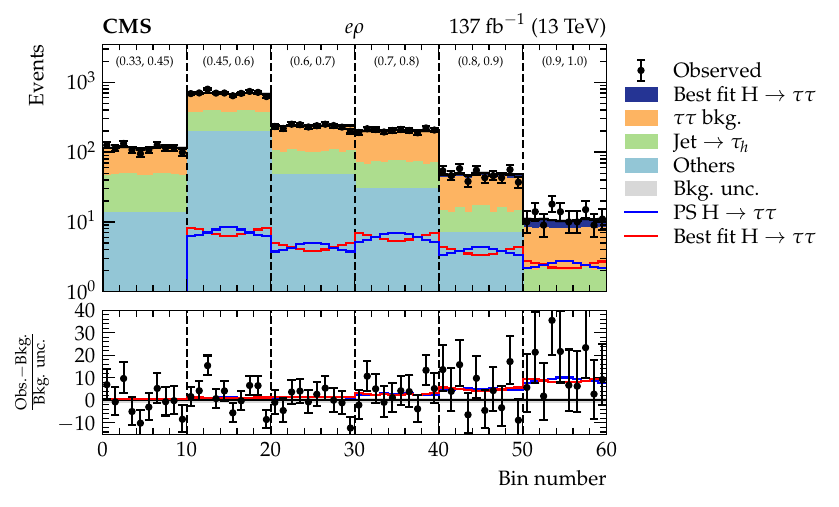}
     \includegraphics[width=1\textwidth]{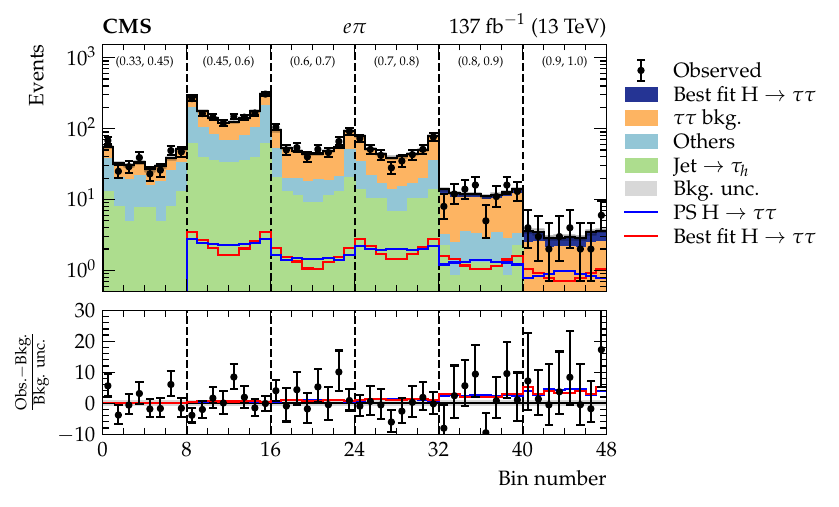}
\caption{Distributions of \phicp in the \erho (upper) and \epi (lower) channels in windows of increasing MVA score, shown on top of each window. The best fit and pseudoscalar (PS) signal distributions are overlaid, where in both cases the signal cross sections are set to the values obtained from the fit to data, which are given in Section~\ref{sec:alpha_results}. The $x$-axis represents the cyclic bins in \phicp in the range of $(0, 360^{\circ})$. In the lower panels, the data minus the background template divided by the uncertainty in the background template is displayed, as well as the signal distributions divided by the uncertainty in the background template. The uncertainty band accounts for all sources of systematic
uncertainty in the background prediction, after the fit to data.\label{fig:unrolledothers}}
\end{figure}

In Figs.~\ref{fig:unrolledmt}--\ref{fig:unrolledothers} we display the post-fit data and background template distributions, after the bin smearing and symmetrisation, with the best fit and pseudoscalar signal templates overlaid. 
The cross sections for the pseudoscalar signal are set to the values determined from the fit to data for the best fit signal, which are given in Section~\ref{sec:alpha_results}.  
The uncertainties have been adjusted to their value after the fit described in Section~\ref{sec:results}. The most sensitive decay modes of the analysis are displayed, which are the \murho and \mupi mode in the \mutau channel displayed in Fig.~\ref{fig:unrolledmt}, the \rhorho and \pirho mode in the \tauhtauh channel displayed in Fig.~\ref{fig:unrolledtt}, and the \erho and \epi channels in the \etau channel displayed in Fig.~\ref{fig:unrolledothers}.
The distributions highlight the effectiveness of the MVA discriminant in optimising the signal over background ratio, as well as the \CP-sensitivity of the measurement that follows from the visibly different phases of the best fit signal and \CP-odd signal distributions. The $180^{\circ}$ phase shift between the \tauhtauh and \ltau channels is manifest in the figures. The correlated effect of the PV smearing is also visible in the \mupi and \epi modes via the non-flat shapes of the background distributions. 

\clearpage

\section{Systematic uncertainties}\label{sec:systematicuncertainties}

The uncertainties considered in this analysis can be categorised into normalisation and shape uncertainties. The former modify only the normalisation of a distribution while leaving its shape unchanged, whereas the latter allow for correlated changes across bins that also alter the shapes of the distributions. The uncertainties are accounted for as nuisance parameters in the fit to data. 
The normalisation and shape uncertainties are summarised in Table~\ref{tab:uncNorm}, in which we also state their correlations between the three different years of data-taking considered in this analysis. 
\begin{table}
\topcaption{Overview of the systematic uncertainties. The third column indicates if the source of uncertainty was treated as being correlated between the years in the fit described in Section~\ref{sec:results}. The fourth column indicates if the uncertainty affects the shapes of the distributions.}
\label{tab:uncNorm}
\centering
\cmsTable{
\begin{tabular}{lccc}
\hline
Uncertainty                     & Magnitude&  Correlation & Shape \\
\hline
Muon reconstruction             & 1\%  &  Yes       & No \\
Electron reconstruction & 2\%  &  Yes         & No \\
Muon trigger                      & 2\% & No     & No  \\
Electron trigger                  & 2\% & No     & No  \\
\etotauh rate in \tauhtauh  & 10 (2)\% 2016 (2017,2018)&  No     & No   \\
\etotauh rate in \etau              & 10\%         &  No     & No  \\
\mutotauh rate in \mutau             & up to 40\% &No      & No  \\
{\PQb}-jet veto                         & 1--9\%&   No    & No  \\
Luminosity                      & 2.3--2.5\%&  Partial   & No \\
Embedded yield                  & 4\%&  No       & No \\
\ttbar cross section            & 4.2\%&   Yes   & No  \\
Diboson cross section       & 5\%&  Yes  & No \\
Single top quark cross section    & 5\%&   Yes  & No \\
\Wjets cross section         & 4\%&   Yes & No  \\
Drell--Yan cross section     & 2\%&   Yes & No \\
\PH cross sections & 2--5\%~\cite{deFlorian:2016spz} &   Yes  & No   \\
\HTT branching fraction & 2\% ~\cite{deFlorian:2016spz} &   Yes & No    \\
SV reco. eff. in \aoneaone &2\%&No & No \\
\tauh ID efficiency &3\% &No & No \\
\SIP In \Pgm, \Pgp, and  \Pe decays  &Decay-mode dependent, 1--5\%& No & No   \\
Muon energy scale            & 0.4--2.7\%& Yes   & Yes     \\
Electron energy scale            & $<$1\%& Yes   & Yes     \\
\tauh Trigger                    & \pt/Decay-mode dependent             & No  & Yes   \\
\tauh Reconstruction             & \pt/Decay-mode dependent (2--3\%)&  Partial    & Yes   \\
Top quark \pt reweighing        &$\pt^{\text{top}}$-Dependent &   Yes      & Yes  \\
\PZ \pt and mass reweighing    & $\pt^{\PZ}/m_{\PZ}$-Dependent &  Partial  & Yes      \\
\tauh Energy scale   & \pt/Decay-mode dependent (0.2--1.1\%) & No          & Yes      \\
\etotauh Energy scale    & 0.5--6.5\% &  No    & Yes  \\
\mutotauh Energy scale   & 1\% &   No  & Yes \\
Jet energy scale                & Event-dependent&  Partial & Yes       \\
Jet energy resolution         & Event-dependent&   No  & Yes  \\
\ptmiss Unclustered scale & Event-dependent&  No  & Yes      \\
\ptmiss Recoil corrections  & Event-dependent&   No & Yes     \\
\FF uncertainties  & Described in text&   Partial & Yes   \\
\ttbar/diboson in embedded          & 10\%&  Yes & Yes       \\
L1 trigger timing (2016--2017)       & Event-dependent (0--4\%)&  Yes & Yes      \\
Renorm./Fact. scales                    & Event-dependent &  Yes  & Yes \\
Parton showering                & Event-dependent &  Yes    & Yes  \\

\hline
\end{tabular}
}
\end{table}

\subsection{Normalisation uncertainties}
The integrated luminosity uncertainty amounts to 2.5, 2.3, and 2.5\% for 2016, 2017, and 2018 respectively~\cite{CMS-PAS-LUM-17-001,CMS-PAS-LUM-17-004,CMS-PAS-LUM-18-002}, and is applied to all simulated samples discussed in Section~\ref{sec:datasimulatedsamples}.

The uncertainty in the muon reconstruction efficiency including the tracking, identification, and isolation requirements is 1\%, while for electrons it is 2\%. The uncertainty in the muon and electron trigger efficiencies, which affect both the single-lepton and cross-triggers, is 2\%. An additional normalisation uncertainty of 4\% is applied to the embedded event samples, originating from the uncertainty in the measurement of the muon trigger and identification efficiencies used to scale the embedded samples.

For the \ltau channels, which contain a veto on events containing {\PQb}-jets, an uncertainty in the propagation of the {\PQb}-quark tagging scale factors of 1--9\% is applied on the \ttbar and diboson event yields (the uncertainties on the event yields for other simulated processes are found to be negligible).

The \FEWZ 3.1 program~\cite{Li_2012} was used to calculate the \Wjets and \Zjets cross sections. Uncertainties in the factorisation and renormalisation scales, the PDF, and the running coupling \alpS were propagated and added in quadrature.
The \TOPPLUSPLUS program~\cite{Czakon_2014} was used to calculate  the \ttbar  cross section and its uncertainty.
The extracted uncertainties for the simulated \Zjets, \Wjets, and \ttbar cross sections amount to 2, 4, and 4\%, respectively.
For the diboson and single top quark production processes, a combined systematic uncertainty in the background yield is estimated to be 5\% using CMS measurements~\cite{Khachatryan_2017,Sirunyan_2017}. The uncertainties in the signal \ggH, VBF, and \VH production cross sections, as well as the uncertainty in the \HTT branching fraction, are applied as recommended in Ref.~\citep{deFlorian:2016spz}.

The uncertainty in the \mutotauh misidentification rate in the \mutau channel is split into four independent uncertainties depending on the MVA decay mode of the \mutotauh candidate. The sizes of the uncertainties are 20\% for \Pgp and \PGr,  30\% for \PaoneONEP, and 40\% for \PaoneTHREEP, respectively.
An uncertainty of 10 (2)\% to the \etotauh misidentification rate is applied for 2016 (2017, 2018) in the \tauhtauh channel. In the \etau channel, the \etotauh misidentification rate is split per decay mode and is, at most, 10\%.

For the \tauhtauh channel, the uncertainty in the \jettotauh background normalisation due to the extrapolation of the \FF from same-sign to opposite-sign regions ranges between 4 and 7\%.

For the decay of the \PGt lepton to \Pgm or a single-pion, an uncertainty in the correction of \SIP is applied by varying the size of the correction by ${\pm}25\%$, while for the decay to an electron the correction is varied by 40\%. The uncertainty is converted into a normalisation uncertainty per decay mode and ranges 1--5\%.
For the \aoneaone mode, the uncertainty in the SV reconstruction efficiency is 2\%. 

Finally, a 3\% uncertainty in the efficiency of the \tauh candidates to pass the DNN discrimination against muon and electron misidentifications is applied.

\subsection{Shape uncertainties}
The uncertainty in the \tauh reconstruction and identification efficiency is typically of the order of 3\%, and split into several uncertainties in each \pt and MVA decay mode bin. The uncertainties in these corrections originates from uncertainties in the fits to the scale factors for these corrections and are statistically dominated. 
We also checked if applying separate uncertainties for \tauh candidates that are incorrectly classified in a different decay mode (\eg \PaoneONEP misclassified as a \PGr) creates any variations in the shapes of the signal or background distributions. However, we found that such uncertainties only resulted in tiny modifications of the shapes of the \phicp distributions, which were negligible in comparison to the statistical uncertainties in the signal and background templates, and therefore common uncertainties were used for correctly and incorrectly classified \tauh candidates in each MVA decay mode bin. 
The uncertainty in the \tauh trigger depends on the \pt and decay mode, and originates from the statistical uncertainty in parameterising the turn-on curve of the triggers.
The \tauh energy scale uncertainty is 0.8--1.1 (0.2--0.5)\% for simulated (embedded) events, and is decay mode dependent. The uncertainty in the \Pgm momentum scale varies as a function of the $\eta$ of the muon and ranges 0.4--2.7\%. The uncertainty in the electron energy scale is less than 1\% and depends on the \pt and $\eta$.
The \etotauh energy scale uncertainty ranges 0.5--6.5\%, while the \mutotauh energy scale uncertainty is 1\%. 

Uncertainties in the jet energy scale originate from different sources with limited correlations. The uncertainties depend on the jet kinematics and are typically larger in the forward regions. Uncertainties in the jet energy resolution are also incorporated; these uncertainties are typically smaller than the jet energy scale uncertainties. 
Uncertainties related to the hadronic recoil response and resolution as derived from the \Zjets,  \Wjets and signal samples, are propagated to \ptvecmiss and observables dependent on \ptvecmiss in the simulated samples that are subject to hadronic recoil corrections. For the samples in which no hadronic recoil is applied (diboson, single top quark, and \ttbar), the jet energy scale and resolution uncertainties as well as the uncertainty in the unclustered energy are propagated to \ptvecmiss and observables dependent on \ptvecmiss in the simulated samples instead.

The embedded samples contain small fractions of \ttbar and diboson events. A shape uncertainty is therefore applied by adding and subtracting 10\% of the simulated \ttbar and diboson contributions. The top quark \pt and Drell--Yan \pt and mass spectra are reweighed. For the top samples, the size of the correction is taken as the uncertainty, while for the Drell--Yan samples the correction is varied by 10\%.

The \FF values are parameterised with continuous functions, and the statistical uncertainties in the fitted parameters are treated as nuisance parameters. 
The uncertainties are parameterised in a manner that allows for asymmetric variations above and below the \pt value where the uncertainty is minimal; the procedure is similar to the method described in detail in Ref.~\cite{Sirunyan_2018}. The size of the correction in \ptmiss is taken as an uncertainty for all \FF values. 
For the \tauhtauh channel, the shape uncertainty in the QCD same-sign to opposite-sign region correction is determined as the difference between a correction binned in the distance \DR~between the two \PGt leptons and the jet multiplicity, and the unbinned correction. 
For the \ltau channels, the equivalent shape uncertainty is taken as the size of the same-sign to opposite-sign correction.
In addition, a systematic uncertainty due to the light-lepton \pt correction is taken as the size of the correction. 
For the \Wjets \FF values, the uncertainty due to the extrapolation from the high-\mT to the low-\mT region is taken as the size of residual differences observed when applying \FF values derived for high-\mT simulated events to low-\mT simulated events. 
For the \ttbar \FF, a systematic uncertainty is applied to account for potential differences between data and simulation. To this purpose, the difference between \FF values derived via data and simulated \Wjets samples is applied as the uncertainty.
An uncertainty in the subtraction of the background processes not involving \jettotauh events is considered by varying the contribution predicted from simulation by ${\pm}$10\%. 

For uncertainties that are common to simulated and embedded samples we treat the lepton and \tauh identification uncertainties and the lepton and  \tauh energy scale uncertainties as being 50\% correlated. All other common uncertainties are treated as being uncorrelated.

During the 2016 and 2017 data-taking periods, a gradual shift in the timing of the inputs of the ECAL L1 trigger in the forward endcap region ($2.5 < \abs{\eta} < 3.0$) led to a specific inefficiency.
Additional correction factors and corresponding uncertainties are applied to the simulation to account for this inefficiency. 
The magnitude of the uncertainties ranges between 0--4\% depending on the process, category, and channel.

For the signal samples, renormalisation and factorisation scales and parton showering uncertainties were incorporated~\cite{deFlorian:2016spz} .

The limited number of events in the signal and background templates is accounted for using the ``Barlow--Beeston" method~\cite{BarlowBeeston,Conway:2011in}, which assigns a single nuisance parameter per bin per process. For background templates that have been flattened as described in Section~\ref{sec:eventcategorisation} the bin-by-bin uncertainties are fully correlated such that there is only one independent nuisance parameter for all \phicp bins. For background templates that are symmetric in $\phicp=180^{\circ}$ one nuisance parameter per pair of symmetrised bins is utilised. It should be noted that for flattened background templates multiple nuisance parameters are still needed per process since multiple windows of increasing MVA score are used.

We also considered other systematic uncertainties that could modify the shape of the simulated \phicp distributions, including the energy scale, and energy and angular resolutions of the charged and neutral pions, impact parameters, and SV$-$PV directions. However, we found that such uncertainties only resulted in tiny modifications to the shapes of the \phicp distributions, which were negligible in comparison to the statistical uncertainties in the signal and background templates, and they were therefore neglected in this analysis.  

The systematic uncertainty scheme is validated by fitting the \phicp distributions in a \ZTT control region, obtained following the procedure described in Section~\ref{ref:validate_model}. Goodness of fit tests have been performed to assess the validity of the statistical model. These tests indicated a good compatibility between the data and the model.

\section{Results}\label{sec:results}

In order to extract the \CP-mixing angle \phitt, a simultaneous fit to the
data is performed using the likelihood function
\likeli that depends on
$\muvec=(\muggh,\muqqh)$, which are the
Higgs boson signal strength modifiers (defined as the cross section times \HTT branching fraction with respect to the SM value), the \CP-mixing angle \phitt,
and the nuisance parameters \vectheta that account for the systematic
uncertainties.
In the fit, all \HTT production processes involving \PV boson couplings, namely VBF and \VH, are scaled by \muqqh, while the \ggH process is scaled by \muggh.
The fit is able to differentiate between these production modes because the shapes of the MVA score distributions are different; the VBF signal tends to peak more sharply towards larger MVA scores, whereas the \ggH distribution is broader. 

The likelihood function is defined as a product of conditional
probabilities $P$ over binned distributions of the discriminating observables in
each event category:
\begin{linenomath}
\begin{equation}
    \likeli = 
    \prod^{N_{\text{categories}}}_{\mathrm{j}} \prod^{N_{\text{bin}}}_{\mathrm{i}} P(\nij \mid \lumi \, \muvec 
   \, \Aij+
  \Bij) \, \prod^{N_{\text{nuisance}}}_{\mathrm{m}} \Cm.
  \label{eq:likelihood}
\end{equation}
\end{linenomath}
In this equation, the Poisson distributions $P$ correspond to the observation of \nij
events in each bin of the discriminating observable given the expectation
for the background \Bij and the signal 
$S_{\text{i,j}}(\lumi,\phitt,\muvec,\vectheta)=\lumi \, \muvec
\, \Aij$, in which \lumi is the integrated
luminosity and \Aij is the signal acceptance in
each production bin.  Constraints on the nuisance parameters corresponding to
the systematic uncertainties described in Section~\ref{sec:systematicuncertainties} are
represented by the functions \Cm. A more detailed discussion on the formulation of the statistical inference may be found in Refs.~\cite{CMS-NOTE-2011-005,Conway:2011in}.
The systematic uncertainties affecting the normalisation of the signal and background templates are incorporated
in the fit via nuisance parameters with a log-normal prior probability density function. The
shape-altering systematic uncertainties are represented by nuisance parameters whose
variations cause continuous morphing of the signal or background template shapes, and are assigned
a Gaussian prior probability density function. The bin-by-bin statistical uncertainties in the background samples are also assigned a Gaussian prior probability density function.

Using the negative log-likelihood, which is defined as
\begin{linenomath}
\begin{equation}
    \DeltaLL = -2 \, \left( \ln(L \phitt) -
    \ln(L \phitt_{\text{best fit}})\right),
    \label{eq:nll}
\end{equation}
\end{linenomath}
we find the 68.3, 95.5, and 99.7\% confidence intervals when \DeltaLL equals 1.00, 4.02, and 8.81, respectively. A detailed discussion may be found in Section 3.2 of Ref.~\cite{Khachatryan:2014jba}.

The inputs to the likelihood fits differ for the signal and background categories. For the signal categories, the \phicp distributions in bins of the MVA score are used (a subset of these are displayed in Figs.~\ref{fig:unrolledmt}--\ref{fig:unrolledothers}). For the background categories, the MVA score distributions are used. This allows for the background contributions and systematic uncertainties to be further constrained, and helps to improve the fit convergence.

\subsection{\texorpdfstring{\phitt}{alpha} mixing angle results}
\label{sec:alpha_results}
\begin{figure}[htb]
    \centering
    \includegraphics[width=0.7\textwidth]{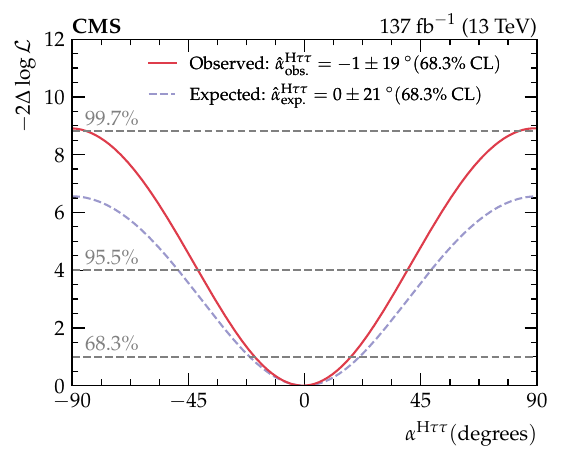}
    \caption{
         Negative log-likelihood scan for the combination of the \etau, \mutau, and \tauhtauh channels.
        The observed (expected) sensitivity to distinguish between the scalar and pseudoscalar hypotheses,
        defined at $\phitt = 0$ and ${\pm} 90^{\circ}$, respectively, is $3.0\sigma$
($2.6\sigma$). The observed (expected) value for \phitt is $-1\pm 19^{\circ}$ ($0\pm 21^{\circ}$) at the 68.3\% \CL. At 95.5\% \CL the range is ${\pm}41^{\circ}$ (${\pm}49^{\circ}$), and at the 99.7\% \CL the observed range is ${\pm}84^{\circ}$.
    \label{fig:LLscan}}
\end{figure}

We present the observed and expected negative log-likelihood scan for the combination of the \etau, \mutau, and \tauhtauh channels in Fig.~\ref{fig:LLscan}. 
The two rate parameters that scale the \ggH and $\Pq\Pq\PH$ production signal strength were left to float freely in the fit. The best fit values of these parameters are $\muggh=0.59^{+0.28}_{-0.32}$ and $\muqqh=1.39^{+0.56}_{-0.47}$, respectively, with the correlation coefficient  $\rho=-0.76$ . We note that there is a strong anticorrelation between these parameters as the analysis does not directly attempt to differentiate between the production modes. 

The data disfavour the pure \CP-odd scenario at $3.0\sigma$. The expected exclusion assuming the SM \PH is $2.6\sigma$. 
 The observed  (expected) value of \phitt is found to be $-1\pm 19^{\circ}$ ($0\pm 21^{\circ}$) at the 68.3\% \CL, and ${\pm} 41^{\circ}$ (${\pm} 49^{\circ}$) at the 95.5\% \CL. 
 Furthermore, we obtain an observed ${\pm} 84^{\circ}$ at the 99.7\% \CL.
The uncertainty can be decomposed into: statistical; bin-by-bin fluctuations in the background templates; experimental systematic uncertainties; and theoretical uncertainties. In this decomposition we obtain 
\begin{linenomath}
\begin{equation*}
\phitt=-1\pm 19\stat\pm 1\syst\pm 2\,\text{(bin-by-bin)} \pm 1\thy^{\circ}.
\end{equation*}
\end{linenomath}
This result is compatible with the SM predictions within the experimental uncertainties.

The expected sensitivities of the \etau, \mutau, and \tauhtauh channels are 1.0, 1.5, and $1.8\sigma$, respectively. 
The \murho mode yields the most sensitive expected contribution of $1.2\sigma$, followed by the \rhorho and \pirho modes that contribute 1.1 and $1.0\sigma$, respectively. All other modes have sensitivities below $1\sigma$.

The statistical uncertainties in the background templates are the subleading source of systematic uncertainty in this analysis. As the dominant contributions to the backgrounds are determined themselves from control samples in data, the amount of data is the limiting factor in this uncertainty.
The next most dominant sources of uncertainty are the hadronic trigger efficiency, theory uncertainties, the \tauh energy scale, and uncertainties related to the implementation of the \FF method.

It was shown in Ref.~\cite{King_2015} that in the next-to-minimal supersymmetric model mixing angles as large as ${\approx}27^{\circ}$ can be accommodated by the latest electric dipole moment and Higgs boson measurements. This measurement is thus sensitive to the larger allowed mixing angles in this model.

\begin{figure}[htb]
  \centering
 \includegraphics[width=0.7\textwidth]{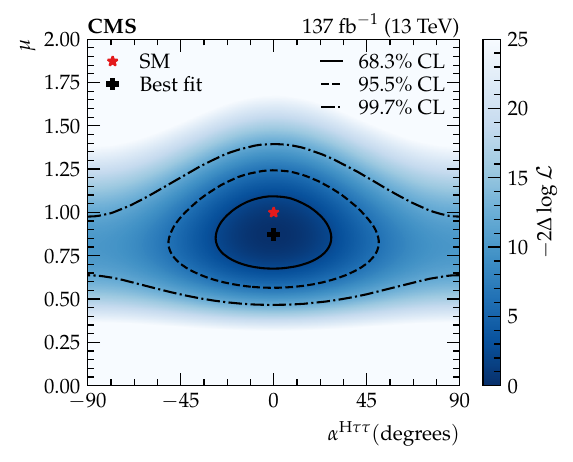} 
  \caption{\label{fig:mu-phitt} The 2-D scan of the signal strength modifier $\mu$ versus \phitt. The 68.3, 95.5, and 99.7\% confidence regions are overlaid.}
\end{figure}

A fit to the data is also performed assuming $\muggh=\muqqh=\mu$. In this case $\mu$ is the combined signal strength modifier that scales the total \PH production cross section times \HTT branching fraction relative to the SM value.
In Fig.~\ref{fig:mu-phitt} we display a scan of $\mu$ versus \phitt.
We observe that there is no strong correlation between these parameters. 

In order to make a 2-D scan of \kappat and \kappattilde, as defined in Eq.~(\ref{eq:phicpinkappa}), we parameterise the likelihood from Eq.~(\ref{eq:likelihood}) in terms of \kappat and \kappattilde. All other \PH couplings are fixed to their expected SM values.

\begin{figure}[htb]
  \centering
 \includegraphics[width=0.7\textwidth]{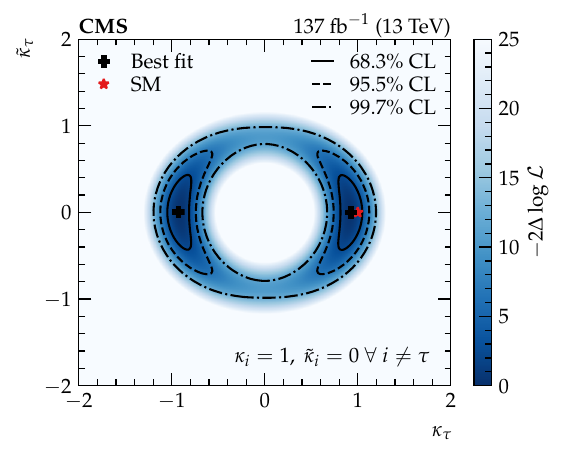}
  \caption{\label{fig:kappa-scan} The 2-D scan of the (reduced) \CP-even (\kappat) and \CP-odd (\kappattilde) \PGt Yukawa couplings. The 68.3, 95.5, and 99.7\% confidence regions are overlaid.}
\end{figure}

In the case of a 2-D negative log-likelihood, the 68.3, 95.5, and 99.7\% confidence
regions are found when $\DeltaLL_{\mathrm{2D}}$ equals 2.30, 6.20, and 11.62~\cite{Khachatryan:2014jba}, respectively, defined analogously to Eq.~(\ref{eq:nll}) with the likelihood now a function of both \kappat and \kappattilde.
All other \CP-even (\CP-odd) couplings affecting the production cross sections and/or the \HTT branching fraction are fixed to their SM values, $\kappa_{i}=1$ ($\widetilde{\kappa}_{i}=0$). 
The observed result of the scan is shown in Fig.~\ref{fig:kappa-scan}.  
It should be noted that the fit is only sensitive to the relative sign between \kappat and \kappattilde and thus the scan has two best fit points for positive and negative values of \kappat.

\begin{figure}[htb]
  \centering
  \includegraphics[width=0.7\textwidth]{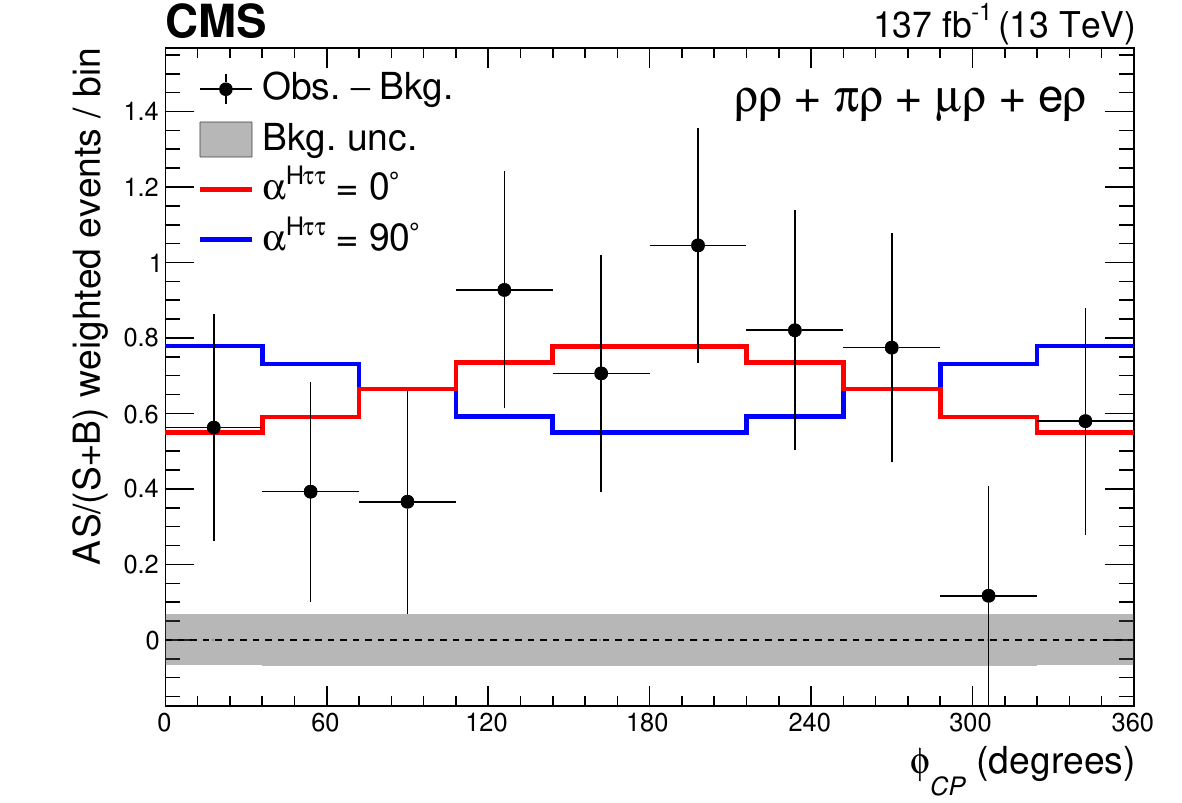}
 \caption{\label{fig:propagandaplot}  The \phicp distributions for the \rhorho, \pirho, \murho, and \erho channels weighed by \AveAsym are combined. Events are included from all MVA score bins in the four signal categories. The background is subtracted from the data. The scalar distribution is depicted in red, while the pseudoscalar is displayed in blue. In the predictions, the rate parameters are taken from their best fit values. The grey uncertainty band indicates the uncertainty in the subtracted background component. In combining the channels, a phase-shift of $180^{\circ}$ was applied to the channels involving a lepton since this channel has a phase difference of $180^{\circ}$ with respect to the two hadronic channels due to a sign-flip in the spectral function of the light lepton.}
\end{figure}

In Fig.~\ref{fig:propagandaplot} we display the data of the \rhorho, \pirho, \murho, and \erho channel together with \CP-even and \CP-odd predictions. 
These channels are chosen as the same number of \phicp bins are used in the fit to data, and collectively they account for most of the sensitivity to \phitt.
Events are included from all MVA score bins in these signal categories. Each MVA score bin is weighed by \AveAsym, where $S$ and $B$ are the signal and background rates, respectively, and $A$ is a measure for the average asymmetry between the scalar and pseudoscalar distributions. The definition of the value of $A$ per bin is ${\abs{\CPeven-\CPodd}/(\CPeven+\CPodd)}$, and $A$ is normalised to the total number of bins. In this equation, \CPeven and \CPodd are the scalar and pseudoscalar contributions per bin.
This distribution shows that the data favour the \CP-even scenario.
 
\clearpage

\section{Summary}\label{sec:summary}

The first measurement of the effective mixing angle \phitt between scalar and pseudoscalar \Yuktau couplings has been presented for a data set of proton-proton collisions at $\sqrt{s}=13\TeV$ corresponding to an integrated luminosity of 137\fbinv. The data were collected with the CMS experiment at the LHC in the period 2016--2018. 
The following \PGt lepton decay modes were included:
 \Pepm, \PGmpm, \Pgppm, \rhotopipiz, $\Paone^\pm\to\Pgppm\Pgpz\Pgpz$, and $\Paone^\pm\to\Pgppm\PGpmp\Pgppm$.
Dedicated strategies were adopted to reconstruct the angle \phicp between the \PGt decay planes for the various \PGt decay modes.
The data disfavour the pure \CP-odd scenario at 3.0 standard deviations.
The observed effective mixing angle is found to be $-1\pm 19^{\circ}$, while the expected value is $0\pm21^{\circ}$ at the 68.3\% confidence level (\CL). 
The observed and expected uncertainties are found to be ${\pm} 41^{\circ}$ and ${\pm}49^{\circ}$ at the 95.5\% \CL, respectively, and the observed sensitivity at the 99.7\% \CL is $\pm84^{\circ}$.
The leading uncertainty in the measurement is statistical, implying that the precision of the measurement will increase with the accumulation of more collision data. 
The measurement is consistent with the standard model expectation, and reduces the allowed parameter space for its extensions.
Tabulated results are provided in HEPDATA~\cite{hepdata}.

\begin{acknowledgments}
  We congratulate our colleagues in the CERN accelerator departments for the excellent performance of the LHC and thank the technical and administrative staffs at CERN and at other CMS institutes for their contributions to the success of the CMS effort. In addition, we gratefully acknowledge the computing centres and personnel of the Worldwide LHC Computing Grid and other centres for delivering so effectively the computing infrastructure essential to our analyses. Finally, we acknowledge the enduring support for the construction and operation of the LHC, the CMS detector, and the supporting computing infrastructure provided by the following funding agencies: BMBWF and FWF (Austria); FNRS and FWO (Belgium); CNPq, CAPES, FAPERJ, FAPERGS, and FAPESP (Brazil); MES and BNSF (Bulgaria); CERN; CAS, MoST, and NSFC (China); MINCIENCIAS (Colombia); MSES and CSF (Croatia); RIF (Cyprus); SENESCYT (Ecuador); MoER, ERC PUT and ERDF (Estonia); Academy of Finland, MEC, and HIP (Finland); CEA and CNRS/IN2P3 (France); BMBF, DFG, and HGF (Germany); GSRI (Greece); NKFIA (Hungary); DAE and DST (India); IPM (Iran); SFI (Ireland); INFN (Italy); MSIP and NRF (Republic of Korea); MES (Latvia); LAS (Lithuania); MOE and UM (Malaysia); BUAP, CINVESTAV, CONACYT, LNS, SEP, and UASLP-FAI (Mexico); MOS (Montenegro); MBIE (New Zealand); PAEC (Pakistan); MSHE and NSC (Poland); FCT (Portugal); JINR (Dubna); MON, RosAtom, RAS, RFBR, and NRC KI (Russia); MESTD (Serbia); SEIDI, CPAN, PCTI, and FEDER (Spain); MOSTR (Sri Lanka); Swiss Funding Agencies (Switzerland); MST (Taipei); ThEPCenter, IPST, STAR, and NSTDA (Thailand); TUBITAK and TAEK (Turkey); NASU (Ukraine); STFC (United Kingdom); DOE and NSF (USA).
  
  \hyphenation{Rachada-pisek} Individuals have received support from the Marie-Curie programme and the European Research Council and Horizon 2020 Grant, contract Nos.\ 675440, 724704, 752730, 758316, 765710, 824093, 884104, and COST Action CA16108 (European Union); the Leventis Foundation; the Alfred P.\ Sloan Foundation; the Alexander von Humboldt Foundation; the Belgian Federal Science Policy Office; the Fonds pour la Formation \`a la Recherche dans l'Industrie et dans l'Agriculture (FRIA-Belgium); the Agentschap voor Innovatie door Wetenschap en Technologie (IWT-Belgium); the F.R.S.-FNRS and FWO (Belgium) under the ``Excellence of Science -- EOS" -- be.h project n.\ 30820817; the Beijing Municipal Science \& Technology Commission, No. Z191100007219010; the Ministry of Education, Youth and Sports (MEYS) of the Czech Republic; the Deutsche Forschungsgemeinschaft (DFG), under Germany's Excellence Strategy -- EXC 2121 ``Quantum Universe" -- 390833306, and under project number 400140256 - GRK2497; the Lend\"ulet (``Momentum") Programme and the J\'anos Bolyai Research Scholarship of the Hungarian Academy of Sciences, the New National Excellence Program \'UNKP, the NKFIA research grants 123842, 123959, 124845, 124850, 125105, 128713, 128786, and 129058 (Hungary); the Council of Science and Industrial Research, India; the Latvian Council of Science; the Ministry of Science and Higher Education and the National Science Center, contracts Opus 2014/15/B/ST2/03998 and 2015/19/B/ST2/02861 (Poland); the Funda\c{c}\~ao para a Ci\^encia e a Tecnologia, grant CEECIND/01334/2018 (Portugal); the National Priorities Research Program by Qatar National Research Fund; the Ministry of Science and Higher Education, projects no. 14.W03.31.0026 and no. FSWW-2020-0008, and the Russian Foundation for Basic Research, project No.19-42-703014 (Russia); the Programa Estatal de Fomento de la Investigaci{\'o}n Cient{\'i}fica y T{\'e}cnica de Excelencia Mar\'{\i}a de Maeztu, grant MDM-2015-0509 and the Programa Severo Ochoa del Principado de Asturias; the Stavros Niarchos Foundation (Greece); the Rachadapisek Sompot Fund for Postdoctoral Fellowship, Chulalongkorn University and the Chulalongkorn Academic into Its 2nd Century Project Advancement Project (Thailand); the Kavli Foundation; the Nvidia Corporation; the SuperMicro Corporation; the Welch Foundation, contract C-1845; and the Weston Havens Foundation (USA).
\end{acknowledgments}

\bibliography{auto_generated}   

\providecommand{\href}[2]{#2}\begingroup\raggedright\begin{thebibliography}{100}%
\makeatletter
\providecommand{\hrefCMSnoop }[0]{\@secondoftwo}%
\makeatother
\providecommand{\doi}{\texttt{doi:}\begingroup \urlstyle{tt}\Url}

\bibitem{PhysRevLett.13.321}
\hrefCMSnoop {}{F.~Englert and R.~Brout, ``Broken symmetry and the mass of
  gauge vector mesons'',} \textit{ Phys. Rev. Lett.} \textbf{ 13} (1964) 321,
  \href{http://dx.doi.org/10.1103/PhysRevLett.13.321}{\doi{10.1103/PhysRevLett.13.321}}.

\bibitem{Higgs:1964ia}
\hrefCMSnoop {}{P.~W. Higgs, ``Broken symmetries, massless particles and gauge
  fields'',} \textit{ Phys. Lett.} \textbf{ 12} (1964) 132,
  \href{http://dx.doi.org/10.1016/0031-9163(64)91136-9}{\doi{10.1016/0031-9163(64)91136-9}}.

\bibitem{PhysRevLett.13.508}
\hrefCMSnoop {}{P.~W. Higgs, ``Broken symmetries and the masses of gauge
  bosons'',} \textit{ Phys. Rev. Lett.} \textbf{ 13} (1964) 508,
  \href{http://dx.doi.org/10.1103/PhysRevLett.13.508}{\doi{10.1103/PhysRevLett.13.508}}.

\bibitem{PhysRevLett.13.585}
\hrefCMSnoop {}{G.~S. Guralnik, C.~R. Hagen, and T.~W.~B. Kibble, ``Global
  conservation laws and massless particles'',} \textit{ Phys. Rev. Lett.}
  \textbf{ 13} (1964) 585,
  \href{http://dx.doi.org/10.1103/PhysRevLett.13.585}{\doi{10.1103/PhysRevLett.13.585}}.

\bibitem{Higgs:1966ev}
\hrefCMSnoop {}{P.~W. Higgs, ``Spontaneous symmetry breakdown without massless
  bosons'',} \textit{ Phys. Rev.} \textbf{ 145} (1966) 1156,
  \href{http://dx.doi.org/10.1103/PhysRev.145.1156}{\doi{10.1103/PhysRev.145.1156}}.

\bibitem{Kibble:1967sv}
\hrefCMSnoop {}{T.~W.~B. Kibble, ``Symmetry breaking in non-abelian gauge
  theories'',} \textit{ Phys. Rev.} \textbf{ 155} (1967) 1554,
  \href{http://dx.doi.org/10.1103/PhysRev.155.1554}{\doi{10.1103/PhysRev.155.1554}}.

\bibitem{Aad_2012}
\hrefCMSnoop {}{{ATLAS Collaboration}, ``Observation of a new particle in the
  search for the standard model {H}iggs boson with the {ATLAS} detector at the
  {LHC}'',} \textit{ Phys. Lett. B} \textbf{ 716} (2012) 1,
  \href{http://dx.doi.org/10.1016/j.physletb.2012.08.020}{\doi{10.1016/j.physletb.2012.08.020}},
  \href{http://www.arXiv.org/abs/1207.7214}{\texttt{arXiv:1207.7214}}.

\bibitem{Chatrchyan_2012}
\hrefCMSnoop {}{{CMS Collaboration}, ``Observation of a new boson at a mass of
  125 {GeV} with the {CMS} experiment at the {LHC}'',} \textit{ Phys. Lett. B}
  \textbf{ 716} (2012) 30,
  \href{http://dx.doi.org/10.1016/j.physletb.2012.08.021}{\doi{10.1016/j.physletb.2012.08.021}},
  \href{http://www.arXiv.org/abs/1207.7235}{\texttt{arXiv:1207.7235}}.

\bibitem{Chatrchyan:2013lba}
\hrefCMSnoop {}{{CMS Collaboration}, ``Observation of a new boson with mass
  near 125 {GeV} in {$\Pp\Pp$} collisions at {$\sqrt{s}$} = 7 and 8 {TeV}'',}
  \textit{ JHEP} \textbf{ 06} (2013) 081,
  \href{http://dx.doi.org/10.1007/JHEP06(2013)081}{\doi{10.1007/JHEP06(2013)081}},
  \href{http://www.arXiv.org/abs/1303.4571}{\texttt{arXiv:1303.4571}}.

\bibitem{Khachatryan:2016vau}
\hrefCMSnoop {}{{ATLAS and CMS Collaborations}, ``Measurements of the {H}iggs
  boson production and decay rates and constraints on its couplings from a
  combined {ATLAS} and {CMS} analysis of the {LHC} pp collision data at {$
  \sqrt{s}=7 $} and 8 {TeV}'',} \textit{ JHEP} \textbf{ 08} (2016) 045,
  \href{http://dx.doi.org/10.1007/JHEP08(2016)045}{\doi{10.1007/JHEP08(2016)045}},
  \href{http://www.arXiv.org/abs/1606.02266}{\texttt{arXiv:1606.02266}}.

\bibitem{Sirunyan_2018}
\hrefCMSnoop {}{{CMS Collaboration}, ``Observation of the {H}iggs boson decay
  to a pair of {$\tau$} leptons with the {CMS} detector'',} \textit{ Phys.
  Lett. B} \textbf{ 779} (2018) 283,
  \href{http://dx.doi.org/10.1016/j.physletb.2018.02.004}{\doi{10.1016/j.physletb.2018.02.004}},
  \href{http://www.arXiv.org/abs/1708.00373}{\texttt{arXiv:1708.00373}}.

\bibitem{Aaboud_2019}
\hrefCMSnoop {}{{ATLAS Collaboration}, ``Cross-section measurements of the
  {H}iggs boson decaying into a pair of {$\tau$}-leptons in proton-proton
  collisions at a centre-of-mass energy of 13 {TeV} with the {ATLAS}
  detector'',} \textit{ Phys. Rev. D} \textbf{ 99} (2019) 072001,
  \href{http://dx.doi.org/10.1103/physrevd.99.072001}{\doi{10.1103/physrevd.99.072001}},
  \href{http://www.arXiv.org/abs/1811.08856}{\texttt{arXiv:1811.08856}}.

\bibitem{Sirunyan:2020xwk}
\hrefCMSnoop {}{{CMS Collaboration}, ``A measurement of the {H}iggs boson mass
  in the diphoton decay channel'',} \textit{ Phys. Lett. B} \textbf{ 805}
  (2020) 135425,
  \href{http://dx.doi.org/10.1016/j.physletb.2020.135425}{\doi{10.1016/j.physletb.2020.135425}},
  \href{http://www.arXiv.org/abs/2002.06398}{\texttt{arXiv:2002.06398}}.

\bibitem{Chatrchyan:2012jja}
\hrefCMSnoop {}{{CMS Collaboration}, ``On the mass and spin-parity of the
  {H}iggs boson candidate via its decays to {Z} boson pairs'',} \textit{ Phys.
  Rev. Lett.} \textbf{ 110} (2013) 081803,
  \href{http://dx.doi.org/10.1103/PhysRevLett.110.081803}{\doi{10.1103/PhysRevLett.110.081803}},
\href{http://www.arXiv.org/abs/1212.6639}{\texttt{arXiv:1212.6639}}.

\bibitem{Chatrchyan:2013mxa}
\hrefCMSnoop {}{{CMS Collaboration}, ``Measurement of the properties of a
  {H}iggs boson in the four-lepton final state'',} \textit{ Phys. Rev. D}
  \textbf{ 89} (2014) 092007,
  \href{http://dx.doi.org/10.1103/PhysRevD.89.092007}{\doi{10.1103/PhysRevD.89.092007}},
\href{http://www.arXiv.org/abs/1312.5353}{\texttt{arXiv:1312.5353}}.

\bibitem{Khachatryan:2014kca}
\hrefCMSnoop {}{{CMS Collaboration}, ``Constraints on the spin-parity and
  anomalous {HVV} couplings of the {H}iggs boson in proton collisions at 7 and
  8 {TeV}'',} \textit{ Phys. Rev. D} \textbf{ 92} (2015) 012004,
  \href{http://dx.doi.org/10.1103/PhysRevD.92.012004}{\doi{10.1103/PhysRevD.92.012004}},
\href{http://www.arXiv.org/abs/1411.3441}{\texttt{arXiv:1411.3441}}.

\bibitem{Khachatryan:2015mma}
\hrefCMSnoop {}{{CMS Collaboration}, ``Limits on the {H}iggs boson lifetime and
  width from its decay to four charged leptons'',} \textit{ Phys. Rev. D}
  \textbf{ 92} (2015) 072010,
  \href{http://dx.doi.org/10.1103/PhysRevD.92.072010}{\doi{10.1103/PhysRevD.92.072010}},
\href{http://www.arXiv.org/abs/1507.06656}{\texttt{arXiv:1507.06656}}.

\bibitem{Khachatryan:2016tnr}
\hrefCMSnoop {}{{CMS Collaboration}, ``Combined search for anomalous
  pseudoscalar {$\PH\PV\PV$} couplings in {$\PV\PH(\PH\to \bbbar$)} production
  and {$\PH\to\PV\PV$} decay'',} \textit{ Phys. Lett. B} \textbf{ 759} (2016)
  672,
  \href{http://dx.doi.org/10.1016/j.physletb.2016.06.004}{\doi{10.1016/j.physletb.2016.06.004}},
\href{http://www.arXiv.org/abs/1602.04305}{\texttt{arXiv:1602.04305}}.

\bibitem{Sirunyan:2017tqd}
\hrefCMSnoop {}{{CMS Collaboration}, ``Constraints on anomalous {H}iggs boson
  couplings using production and decay information in the four-lepton final
  state'',} \textit{ Phys. Lett. B} \textbf{ 775} (2017) 1,
  \href{http://dx.doi.org/10.1016/j.physletb.2017.10.021}{\doi{10.1016/j.physletb.2017.10.021}},
\href{http://www.arXiv.org/abs/1707.00541}{\texttt{arXiv:1707.00541}}.

\bibitem{Sirunyan_2019}
\hrefCMSnoop {}{{{CMS}} Collaboration, ``Constraints on anomalous {HVV}
  couplings from the production of {H}iggs bosons decaying to {$\tau$} lepton
  pairs'',} \textit{ Phys. Rev. D} \textbf{ 100} (2019) 112002,
  \href{http://dx.doi.org/10.1103/physrevd.100.112002}{\doi{10.1103/physrevd.100.112002}},
  \href{http://www.arXiv.org/abs/1903.06973}{\texttt{arXiv:1903.06973}}.

\bibitem{CMS:2021nnc}
\hrefCMSnoop {}{{CMS Collaboration}, ``Constraints on anomalous {H}iggs boson
  couplings to vector bosons and fermions in its production and decay using the
  four-lepton final state'',} \textit{ Phys. Rev. D} \textbf{ 104} (2021)
  052004,
  \href{http://dx.doi.org/10.1103/PhysRevD.104.052004}{\doi{10.1103/PhysRevD.104.052004}},
  \href{http://www.arXiv.org/abs/2104.12152}{\texttt{arXiv:2104.12152}}.

\bibitem{Aad:2013xqa}
\hrefCMSnoop {}{{ATLAS Collaboration}, ``Evidence for the spin-0 nature of the
  {H}iggs boson using {ATLAS} data'',} \textit{ Phys. Lett. B} \textbf{ 726}
  (2013) 120,
  \href{http://dx.doi.org/10.1016/j.physletb.2013.08.026}{\doi{10.1016/j.physletb.2013.08.026}},
\href{http://www.arXiv.org/abs/1307.1432}{\texttt{arXiv:1307.1432}}.

\bibitem{Aad:2015mxa}
\hrefCMSnoop {}{{ATLAS Collaboration}, ``Study of the spin and parity of the
  {H}iggs boson in diboson decays with the {ATLAS} detector'',} \textit{ Eur.
  Phys. J. C} \textbf{ 75} (2015) 476,
  \href{http://dx.doi.org/10.1140/epjc/s10052-015-3685-1}{\doi{10.1140/epjc/s10052-015-3685-1}},
  \href{http://www.arXiv.org/abs/1506.05669}{\texttt{arXiv:1506.05669}}.
  [Erratum: \DOI{10.1140/epjc/s10052-016-3934-y}].

\bibitem{Aad:2016nal}
\hrefCMSnoop {}{{ATLAS Collaboration}, ``Test of {$CP$} invariance in
  vector-boson fusion production of the {H}iggs boson using the optimal
  observable method in the ditau decay channel with the {ATLAS} detector'',}
  \textit{ Eur. Phys. J. C} \textbf{ 76} (2016) 658,
  \href{http://dx.doi.org/10.1140/epjc/s10052-016-4499-5}{\doi{10.1140/epjc/s10052-016-4499-5}},
\href{http://www.arXiv.org/abs/1602.04516}{\texttt{arXiv:1602.04516}}.

\bibitem{Aaboud:2017oem}
\hrefCMSnoop {}{{ATLAS Collaboration}, ``Measurement of inclusive and
  differential cross sections in the {$\PH \to \PZ\PZ^* \to 4\ell$} decay
  channel in {$\Pp\Pp$} collisions at {$\sqrt{s}=13$} {TeV} with the {ATLAS}
  detector'',} \textit{ JHEP} \textbf{ 10} (2017) 132,
  \href{http://dx.doi.org/10.1007/JHEP10(2017)132}{\doi{10.1007/JHEP10(2017)132}},
  \href{http://www.arXiv.org/abs/1708.02810}{\texttt{arXiv:1708.02810}}.

\bibitem{Aaboud:2017vzb}
\hrefCMSnoop {}{{ATLAS Collaboration}, ``Measurement of the {H}iggs boson
  coupling properties in the {$\PH \to \PZ\PZ^* \to 4\ell$} decay channel at
  {$\sqrt{s} = 13\TeV$} with the {ATLAS} detector'',} \textit{ JHEP} \textbf{
  03} (2018) 095,
  \href{http://dx.doi.org/10.1007/JHEP03(2018)095}{\doi{10.1007/JHEP03(2018)095}},
\href{http://www.arXiv.org/abs/1712.02304}{\texttt{arXiv:1712.02304}}.

\bibitem{Aaboud:2018xdt}
\hrefCMSnoop {}{{ATLAS Collaboration}, ``Measurements of {H}iggs boson
  properties in the diphoton decay channel with 36 {\fbinv} of {\PP} collision
  data at {$\sqrt{s} = 13\TeV$} with the {ATLAS} detector'',} \textit{ Phys.
  Rev. D} \textbf{ 98} (2018) 052005,
  \href{http://dx.doi.org/10.1103/PhysRevD.98.052005}{\doi{10.1103/PhysRevD.98.052005}},
\href{http://www.arXiv.org/abs/1802.04146}{\texttt{arXiv:1802.04146}}.

\bibitem{Zhang_2011}
\hrefCMSnoop {}{C.~Zhang and S.~Willenbrock, ``Effective-field-theory approach
  to top-quark production and decay'',} \textit{ Phys. Rev. D} \textbf{ 83}
  (2011) 034006,
  \href{http://dx.doi.org/10.1103/physrevd.83.034006}{\doi{10.1103/physrevd.83.034006}},
  \href{http://www.arXiv.org/abs/1008.3869}{\texttt{arXiv:1008.3869}}.

\bibitem{Harnik:2013aja}
R.~Harnik\hrefCMSnoop {}{ {et~al.}, ``Measuring {$CP$} violation in {$\PH \to
  \tau^+ \tau^-$} at colliders'',} \textit{ Phys. Rev. D} \textbf{ 88} (2013)
  076009,
  \href{http://dx.doi.org/10.1103/PhysRevD.88.076009}{\doi{10.1103/PhysRevD.88.076009}},
  \href{http://www.arXiv.org/abs/1308.1094}{\texttt{arXiv:1308.1094}}.

\bibitem{Ghosh_2019}
\hrefCMSnoop {}{T.~Ghosh, R.~Godbole, and X.~Tata, ``Determining the spacetime
  structure of bottom-quark couplings to spin-zero particles'',} \textit{ Phys.
  Rev. D} \textbf{ 100} (2019) 015026,
  \href{http://dx.doi.org/10.1103/physrevd.100.015026}{\doi{10.1103/physrevd.100.015026}},
  \href{http://www.arXiv.org/abs/1904.09895}{\texttt{arXiv:1904.09895}}.

\bibitem{Gritsan_2016}
\hrefCMSnoop {}{A.~V. Gritsan, R.~R{\"o}ntsch, M.~Schulze, and M.~Xiao,
  ``Constraining anomalous {H}iggs boson couplings to the heavy-flavor fermions
  using matrix element techniques'',} \textit{ Phys. Rev. D} \textbf{ 94}
  (2016) 055023,
  \href{http://dx.doi.org/10.1103/physrevd.94.055023}{\doi{10.1103/physrevd.94.055023}},
  \href{http://www.arXiv.org/abs/1606.03107}{\texttt{arXiv:1606.03107}}.

\bibitem{CMSTOPCP2020}
\hrefCMSnoop {}{{CMS Collaboration}, ``Measurements of {$\ttbar\PH$} production
  and the {$CP$} structure of the {Y}ukawa interaction between the {H}iggs
  boson and top quark in the diphoton decay channel'',} \textit{ Phys. Rev.
  Lett.} \textbf{ 125} (2020) 061801,
  \href{http://dx.doi.org/10.1103/PhysRevLett.125.061801}{\doi{10.1103/PhysRevLett.125.061801}},
  \href{http://www.arXiv.org/abs/2003.10866}{\texttt{arXiv:2003.10866}}.

\bibitem{ATLASTOPCP2020}
\hrefCMSnoop {}{{ATLAS Collaboration}, ``{$CP$} properties of {H}iggs boson
  interactions with top quarks in the {$\ttbar\PH$} and {$\PQt\PH$} processes
  using {$\PH \to \PGg\PGg$} with the {ATLAS} detector'',} \textit{ Phys. Rev.
  Lett.} \textbf{ 125} (2020) 061802,
  \href{http://dx.doi.org/10.1103/PhysRevLett.125.061802}{\doi{10.1103/PhysRevLett.125.061802}},
  \href{http://www.arXiv.org/abs/2004.04545}{\texttt{arXiv:2004.04545}}.

\bibitem{ATLAS:2021pkb}
\hrefCMSnoop {}{{ATLAS Collaboration}, ``Constraints on {H}iggs boson
  properties using {$\PW\PW^{*}(\rightarrow \Pe\nu\mu\nu) jj$} production in
  36.1{\fbinv} of {$\sqrt{s}$}=13 {TeV} pp collisions with the {ATLAS}
  detector'',} 2021.
  \href{http://www.arXiv.org/abs/2109.13808}{\texttt{arXiv:2109.13808}}.
  Accepted by \textit{Eur. Phys. J. C}.

\bibitem{Fontes:2015mea}
\hrefCMSnoop {}{D.~Fontes, J.~C. Rom$\tilde{\text{a}}$o, R.~Santos, and J.~P.
  Silva, ``Large pseudoscalar {Y}ukawa couplings in the complex {2HDM}'',}
  \textit{ JHEP} \textbf{ 06} (2015) 060,
  \href{http://dx.doi.org/10.1007/JHEP06(2015)060}{\doi{10.1007/JHEP06(2015)060}},
\href{http://www.arXiv.org/abs/1502.01720}{\texttt{arXiv:1502.01720}}.

\bibitem{King_2015}
\hrefCMSnoop {}{S.~F. King, M.~M{\"u}hlleitner, R.~Nevzorov, and K.~Walz,
  ``Exploring the {$CP$}-violating {NMSSM}: {EDM} constraints and
  phenomenology'',} \textit{ Nucl. Phys. B} \textbf{ 901} (2015) 526,
  \href{http://dx.doi.org/10.1016/j.nuclphysb.2015.11.003}{\doi{10.1016/j.nuclphysb.2015.11.003}},
  \href{http://www.arXiv.org/abs/1508.03255}{\texttt{arXiv:1508.03255}}.

\bibitem{Berge2014}
\hrefCMSnoop {}{S.~Berge, W.~Bernreuther, and S.~Kirchner, ``Determination of
  the {H}iggs {$CP$}-mixing angle in the tau decay channels at the {LHC}
  including the {D}rell--{Y}an background'',} \textit{ Eur. Phys. J. C}
  \textbf{ 74} (2014) 3164,
  \href{http://dx.doi.org/10.1140/epjc/s10052-014-3164-0}{\doi{10.1140/epjc/s10052-014-3164-0}},
  \href{http://www.arXiv.org/abs/1408.0798}{\texttt{arXiv:1408.0798}}.

\bibitem{Zyla:2020zbs}
\hrefCMSnoop {}{{Particle Data Group}, P.~A. Zyla {et~al.}, ``Review of
  particle physics'',} \textit{ Prog. Theor. Exp. Phys.} \textbf{ 2020} (2020)
  083C01,
  \href{http://dx.doi.org/10.1093/ptep/ptaa104}{\doi{10.1093/ptep/ptaa104}}.

\bibitem{BERGE2016841}
\hrefCMSnoop {}{S.~Berge, W.~Bernreuther, and S.~Kirchner, ``Determination of
  the {H}iggs {$CP$}-mixing angle in the tau decay channels'',} \textit{ Nucl.
  Part. Phys. Proc.} \textbf{ 273-275} (2016) 841,
  \href{http://dx.doi.org/10.1016/j.nuclphysbps.2015.09.129}{\doi{10.1016/j.nuclphysbps.2015.09.129}},
  \href{http://www.arXiv.org/abs/1410.6362}{\texttt{arXiv:1410.6362}}.

\bibitem{Berge:2011ij}
\hrefCMSnoop {}{S.~Berge, W.~Bernreuther, B.~Niepelt, and H.~Spiesberger, ``How
  to pin down the {$CP$} quantum numbers of a {H}iggs boson in its tau decays
  at the {LHC}'',} \textit{ Phys. Rev. D} \textbf{ 84} (2011) 116003,
  \href{http://dx.doi.org/10.1103/PhysRevD.84.116003}{\doi{10.1103/PhysRevD.84.116003}},
  \href{http://www.arXiv.org/abs/1108.0670}{\texttt{arXiv:1108.0670}}.

\bibitem{PhysRevD.33.93}
\hrefCMSnoop {}{J.~R. Dell'Aquila and C.~A. Nelson, ``Distinguishing a spin-0
  technipion and an elementary {H}iggs boson: ${V}_{1}$${V}_{2}$ modes with
  decays into $\textit{I}^-_{A} \boldsymbol{l}_{B}$ and/or $q^-_{A}
  \boldsymbol{q}_{B}$'',} \textit{ Phys. Rev. D} \textbf{ 33} (1986) 93,
  \href{http://dx.doi.org/10.1103/PhysRevD.33.93}{\doi{10.1103/PhysRevD.33.93}}.

\bibitem{Kr_mer_1994}
\hrefCMSnoop {}{M.~Kr{\"a}mer, J.~K{\"u}hn, M.~L. Stong, and P.~M. Zerwas,
  ``Prospects of measuring the parity of {H}iggs particles'',} \textit{ Z.
  Phys. C} \textbf{ 64} (1994) 21,
  \href{http://dx.doi.org/10.1007/bf01557231}{\doi{10.1007/bf01557231}},
  \href{http://www.arXiv.org/abs/hep-ph/9404280}{\texttt{arXiv:hep-ph/9404280}}.

\bibitem{Sirunyan:2020zal}
\hrefCMSnoop {}{{CMS Collaboration}, ``Performance of the {CMS} {L}evel-1
  trigger in proton-proton collisions at {$\sqrt{s} = 13\TeV$}'',} \textit{
  JINST} \textbf{ 15} (2020) P10017,
  \href{http://dx.doi.org/10.1088/1748-0221/15/10/P10017}{\doi{10.1088/1748-0221/15/10/P10017}},
  \href{http://www.arXiv.org/abs/2006.10165}{\texttt{arXiv:2006.10165}}.

\bibitem{Khachatryan:2016bia}
\hrefCMSnoop {}{{CMS Collaboration}, ``The {CMS} trigger system'',} \textit{
  JINST} \textbf{ 12} (2017) P01020,
  \href{http://dx.doi.org/10.1088/1748-0221/12/01/P01020}{\doi{10.1088/1748-0221/12/01/P01020}},
\href{http://www.arXiv.org/abs/1609.02366}{\texttt{arXiv:1609.02366}}.

\bibitem{Chatrchyan:2008zzk}
\hrefCMSnoop {}{{CMS Collaboration}, ``The {CMS} experiment at the {CERN}
  {LHC}'',} \textit{ JINST} \textbf{ 3} (2008) S08004,
  \href{http://dx.doi.org/10.1088/1748-0221/3/08/S08004}{\doi{10.1088/1748-0221/3/08/S08004}}.

\bibitem{Nason:2004rx}
\hrefCMSnoop {}{P.~Nason, ``A new method for combining {NLO} {QCD} with shower
  {M}onte {C}arlo algorithms'',} \textit{ JHEP} \textbf{ 11} (2004) 040,
  \href{http://dx.doi.org/10.1088/1126-6708/2004/11/040}{\doi{10.1088/1126-6708/2004/11/040}},
\href{http://www.arXiv.org/abs/hep-ph/0409146}{\texttt{arXiv:hep-ph/0409146}}.

\bibitem{Frixione:2007vw}
\hrefCMSnoop {}{S.~Frixione, P.~Nason, and C.~Oleari, ``Matching {NLO} {QCD}
  computations with parton shower simulations: the {POWHEG} method'',} \textit{
  JHEP} \textbf{ 11} (2007) 070,
  \href{http://dx.doi.org/10.1088/1126-6708/2007/11/070}{\doi{10.1088/1126-6708/2007/11/070}},
\href{http://www.arXiv.org/abs/0709.2092}{\texttt{arXiv:0709.2092}}.

\bibitem{Alioli:2010xd}
\hrefCMSnoop {}{S.~Alioli, P.~Nason, C.~Oleari, and E.~Re, ``A general
  framework for implementing {NLO} calculations in shower {M}onte {C}arlo
  programs: the {POWHEG BOX}'',} \textit{ JHEP} \textbf{ 06} (2010) 043,
  \href{http://dx.doi.org/10.1007/JHEP06(2010)043}{\doi{10.1007/JHEP06(2010)043}},
\href{http://www.arXiv.org/abs/1002.2581}{\texttt{arXiv:1002.2581}}.

\bibitem{Bagnaschi:2011tu}
\hrefCMSnoop {}{E.~Bagnaschi, G.~Degrassi, P.~Slavich, and A.~Vicini, ``{H}iggs
  production via gluon fusion in the {POWHEG} approach in the {SM} and in the
  {MSSM}'',} \textit{ JHEP} \textbf{ 02} (2012) 088,
  \href{http://dx.doi.org/10.1007/JHEP02(2012)088}{\doi{10.1007/JHEP02(2012)088}},
  \href{http://www.arXiv.org/abs/1111.2854}{\texttt{arXiv:1111.2854}}.

\bibitem{Nason:2009ai}
\hrefCMSnoop {}{P.~Nason and C.~Oleari, ``{NLO} {H}iggs boson production via
  vector-boson fusion matched with shower in {POWHEG}'',} \textit{ JHEP}
  \textbf{ 02} (2010) 037,
  \href{http://dx.doi.org/10.1007/JHEP02(2010)037}{\doi{10.1007/JHEP02(2010)037}},
  \href{http://www.arXiv.org/abs/0911.5299}{\texttt{arXiv:0911.5299}}.

\bibitem{Jezo:2015aia}
\hrefCMSnoop {}{T.~Je{\v{z}}o and P.~Nason, ``On the treatment of resonances in
  next-to-leading order calculations matched to a parton shower'',} \textit{
  JHEP} \textbf{ 12} (2015) 065,
  \href{http://dx.doi.org/10.1007/JHEP12(2015)065}{\doi{10.1007/JHEP12(2015)065}},
  \href{http://www.arXiv.org/abs/1509.09071}{\texttt{arXiv:1509.09071}}.

\bibitem{Granata:2017iod}
\hrefCMSnoop {}{F.~Granata, J.~M. Lindert, C.~Oleari, and S.~Pozzorini, ``{NLO}
  {QCD+EW} predictions for {HV} and {HV}+jet production including parton-shower
  effects'',} \textit{ JHEP} \textbf{ 09} (2017) 012,
  \href{http://dx.doi.org/10.1007/JHEP09(2017)012}{\doi{10.1007/JHEP09(2017)012}},
  \href{http://www.arXiv.org/abs/1706.03522}{\texttt{arXiv:1706.03522}}.

\bibitem{Klamke:2007cu}
\hrefCMSnoop {}{G.~Klamke and D.~Zeppenfeld, ``{H}iggs plus two jet production
  via gluon fusion as a signal at the {CERN LHC}'',} \textit{ JHEP} \textbf{
  04} (2007) 052,
  \href{http://dx.doi.org/10.1088/1126-6708/2007/04/052}{\doi{10.1088/1126-6708/2007/04/052}},
  \href{http://www.arXiv.org/abs/hep-ph/0703202}{\texttt{arXiv:hep-ph/0703202}}.

\bibitem{Demartin:2014fia}
F.~Demartin\hrefCMSnoop {}{ {et~al.}, ``{Higgs} characterisation at {NLO} in
  {QCD}: {$CP$} properties of the top-quark {Y}ukawa interaction'',} \textit{
  Eur. Phys. J. C} \textbf{ 74} (2014) 3065,
  \href{http://dx.doi.org/10.1140/epjc/s10052-014-3065-2}{\doi{10.1140/epjc/s10052-014-3065-2}},
\href{http://www.arXiv.org/abs/1407.5089}{\texttt{arXiv:1407.5089}}.

\bibitem{Hamilton:2013fea}
\hrefCMSnoop {}{K.~Hamilton, P.~Nason, E.~Re, and G.~Zanderighi, ``{NNLOPS}
  simulation of {H}iggs boson production'',} \textit{ JHEP} \textbf{ 10} (2013)
  222,
  \href{http://dx.doi.org/10.1007/JHEP10(2013)222}{\doi{10.1007/JHEP10(2013)222}},
  \href{http://www.arXiv.org/abs/1309.0017}{\texttt{arXiv:1309.0017}}.

\bibitem{Hamilton:2015nsa}
\hrefCMSnoop {}{K.~Hamilton, P.~Nason, and G.~Zanderighi, ``Finite quark-mass
  effects in the {NNLOPS POWHEG+MiNLO} {H}iggs generator'',} \textit{ JHEP}
  \textbf{ 05} (2015) 140,
  \href{http://dx.doi.org/10.1007/JHEP05(2015)140}{\doi{10.1007/JHEP05(2015)140}},
  \href{http://www.arXiv.org/abs/1501.04637}{\texttt{arXiv:1501.04637}}.

\bibitem{Sjostrand:2014zea}
T.~Sj{\"o}strand\hrefCMSnoop {}{ {et~al.}, ``An introduction to {PYTHIA}
  8.2'',} \textit{ Comput. Phys. Commun.} \textbf{ 191} (2015) 159,
  \href{http://dx.doi.org/10.1016/j.cpc.2015.01.024}{\doi{10.1016/j.cpc.2015.01.024}},
\href{http://www.arXiv.org/abs/1410.3012}{\texttt{arXiv:1410.3012}}.

\bibitem{Przedzinski:2018ett}
\hrefCMSnoop {}{T.~Przedzinski, E.~Richter-Was, and Z.~Was, ``Documentation of
  {T}au{S}pinner algorithms: program for simulating spin effects in
  {$\tau$}-lepton production at {LHC}'',} \textit{ Eur. Phys. J. C} \textbf{
  79} (2019) 91,
  \href{http://dx.doi.org/10.1140/epjc/s10052-018-6527-0}{\doi{10.1140/epjc/s10052-018-6527-0}},
\href{http://www.arXiv.org/abs/1802.05459}{\texttt{arXiv:1802.05459}}.

\bibitem{Ball_2015}
\hrefCMSnoop {}{{NNPDF} Collaboration, ``Parton distributions for the {LHC}
  {R}un {II}'',} \textit{ JHEP} \textbf{ 04} (2015) 040,
  \href{http://dx.doi.org/10.1007/JHEP04(2015)040}{\doi{10.1007/JHEP04(2015)040}},
  \href{http://www.arXiv.org/abs/1410.8849}{\texttt{arXiv:1410.8849}}.

\bibitem{Ball_2017}
\hrefCMSnoop {}{{NNPDF} Collaboration, ``Parton distributions from
  high-precision collider data'',} \textit{ Eur. Phys. J. C} \textbf{ 77}
  (2017) 663,
  \href{http://dx.doi.org/10.1140/epjc/s10052-017-5199-5}{\doi{10.1140/epjc/s10052-017-5199-5}},
  \href{http://www.arXiv.org/abs/1706.00428}{\texttt{arXiv:1706.00428}}.

\bibitem{Alwall:2014hca}
J.~Alwall\hrefCMSnoop {}{ {et~al.}, ``The automated computation of tree-level
  and next-to-leading order differential cross sections, and their matching to
  parton shower simulations'',} \textit{ JHEP} \textbf{ 07} (2014) 079,
  \href{http://dx.doi.org/10.1007/JHEP07(2014)079}{\doi{10.1007/JHEP07(2014)079}},
\href{http://www.arXiv.org/abs/1405.0301}{\texttt{arXiv:1405.0301}}.

\bibitem{Alwall:2007fs}
J.~Alwall\hrefCMSnoop {}{ {et~al.}, ``Comparative study of various algorithms
  for the merging of parton showers and matrix elements in hadronic
  collisions'',} \textit{ Eur. Phys. J. C} \textbf{ 53} (2008) 473,
  \href{http://dx.doi.org/10.1140/epjc/s10052-007-0490-5}{\doi{10.1140/epjc/s10052-007-0490-5}},
\href{http://www.arXiv.org/abs/0706.2569}{\texttt{arXiv:0706.2569}}.

\bibitem{Alioli:2011as}
\hrefCMSnoop {}{S.~Alioli, S.-O. Moch, and P.~Uwer, ``Hadronic top-quark
  pair-production with one jet and parton showering'',} \textit{ JHEP} \textbf{
  01} (2012) 137,
  \href{http://dx.doi.org/10.1007/JHEP01(2012)137}{\doi{10.1007/JHEP01(2012)137}},
  \href{http://www.arXiv.org/abs/1110.5251}{\texttt{arXiv:1110.5251}}.

\bibitem{Re:2010bp}
\hrefCMSnoop {}{E.~Re, ``Single-top {$Wt$}-channel production matched with
  parton showers using the {POWHEG} method'',} \textit{ Eur. Phys. J. C}
  \textbf{ 71} (2011) 1547,
  \href{http://dx.doi.org/10.1140/epjc/s10052-011-1547-z}{\doi{10.1140/epjc/s10052-011-1547-z}},
  \href{http://www.arXiv.org/abs/1009.2450}{\texttt{arXiv:1009.2450}}.

\bibitem{Frederix:2012dh}
\hrefCMSnoop {}{R.~Frederix, E.~Re, and P.~Torrielli, ``Single-top
  {$t$}-channel hadroproduction in the four-flavour scheme with {POWHEG} and
  {aMC@NLO}'',} \textit{ JHEP} \textbf{ 09} (2012) 130,
  \href{http://dx.doi.org/10.1007/JHEP09(2012)130}{\doi{10.1007/JHEP09(2012)130}},
  \href{http://www.arXiv.org/abs/1207.5391}{\texttt{arXiv:1207.5391}}.

\bibitem{Khachatryan:2015pea}
\hrefCMSnoop {}{{CMS Collaboration}, ``Event generator tunes obtained from
  underlying event and multiparton scattering measurements'',} \textit{ Eur.
  Phys. J. C} \textbf{ 76} (2016) 155,
  \href{http://dx.doi.org/10.1140/epjc/s10052-016-3988-x}{\doi{10.1140/epjc/s10052-016-3988-x}},
\href{http://www.arXiv.org/abs/1512.00815}{\texttt{arXiv:1512.00815}}.

\bibitem{Sirunyan_2020_CP5}
\hrefCMSnoop {}{{CMS Collaboration}, ``Extraction and validation of a new set
  of {CMS PYTHIA8} tunes from underlying-event measurements'',} \textit{ Eur.
  Phys. J. C} \textbf{ 80} (2020) 4,
  \href{http://dx.doi.org/10.1140/epjc/s10052-019-7499-4}{\doi{10.1140/epjc/s10052-019-7499-4}},
  \href{http://www.arXiv.org/abs/1903.12179}{\texttt{arXiv:1903.12179}}.

\bibitem{Agostinelli:2002hh}
\hrefCMSnoop {}{{GEANT4} Collaboration, ``{\GEANTfour}---a simulation
  toolkit'',} \textit{ Nucl. Instrum. Meth. A} \textbf{ 506} (2003) 250,
\href{http://dx.doi.org/10.1016/S0168-9002(03)01368-8}{\doi{10.1016/S0168-9002(03)01368-8}}.

\bibitem{Sirunyan:2017ulk}
\hrefCMSnoop {}{{CMS Collaboration}, ``Particle-flow reconstruction and global
  event description with the {CMS} detector'',} \textit{ JINST} \textbf{ 12}
  (2017) P10003,
  \href{http://dx.doi.org/10.1088/1748-0221/12/10/P10003}{\doi{10.1088/1748-0221/12/10/P10003}},
\href{http://www.arXiv.org/abs/1706.04965}{\texttt{arXiv:1706.04965}}.

\bibitem{TRK-11-001}
\hrefCMSnoop {}{{CMS Collaboration}, ``Description and performance of track and
  primary-vertex reconstruction with the {CMS} tracker'',} \textit{ JINST}
  \textbf{ 9} (2014) P10009,
  \href{http://dx.doi.org/10.1088/1748-0221/9/10/P10009}{\doi{10.1088/1748-0221/9/10/P10009}},
\href{http://www.arXiv.org/abs/1405.6569}{\texttt{arXiv:1405.6569}}.

\bibitem{726788}
\hrefCMSnoop {}{K.~Rose, ``Deterministic annealing for clustering, compression,
  classification, regression, and related optimization problems'',} \textit{
  Proc. IEEE} \textbf{ 86} (1998) 2210,
  \href{http://dx.doi.org/10.1109/5.726788}{\doi{10.1109/5.726788}}.

\bibitem{Waltenberger_2007}
\hrefCMSnoop {}{W.~Waltenberger, R.~Fr{\"u}hwirth, and P.~Vanlaer, ``Adaptive
  vertex fitting'',} \textit{ J. Phys. G: Nuc. Part. Phys.} \textbf{ 34} (2007)
  N343,
  \href{http://dx.doi.org/10.1088/0954-3899/34/12/n01}{\doi{10.1088/0954-3899/34/12/n01}}.

\bibitem{Cacciari:2008gp}
\hrefCMSnoop {}{M.~Cacciari, G.~P. Salam, and G.~Soyez, ``The anti-\kt jet
  clustering algorithm'',} \textit{ JHEP} \textbf{ 04} (2008) 063,
  \href{http://dx.doi.org/10.1088/1126-6708/2008/04/063}{\doi{10.1088/1126-6708/2008/04/063}},
  \href{http://www.arXiv.org/abs/0802.1189}{\texttt{arXiv:0802.1189}}.

\bibitem{Chatrchyan:2012xi}
\hrefCMSnoop {}{{CMS Collaboration}, ``Performance of {CMS} muon reconstruction
  in pp collision events at {$\sqrt{s}=7$} {TeV}'',} \textit{ JINST} \textbf{
  7} (2012) P10002,
  \href{http://dx.doi.org/10.1088/1748-0221/7/10/P10002}{\doi{10.1088/1748-0221/7/10/P10002}},
\href{http://www.arXiv.org/abs/1206.4071}{\texttt{arXiv:1206.4071}}.

\bibitem{Khachatryan:2015hwa}
\hrefCMSnoop {}{{CMS Collaboration}, ``Performance of electron reconstruction
  and selection with the {CMS} detector in proton-proton collisions at
  {$\sqrt{s} = 8$} {TeV}'',} \textit{ JINST} \textbf{ 10} (2015) P06005,
  \href{http://dx.doi.org/10.1088/1748-0221/10/06/P06005}{\doi{10.1088/1748-0221/10/06/P06005}},
\href{http://www.arXiv.org/abs/1502.02701}{\texttt{arXiv:1502.02701}}.

\bibitem{Cacciari:2011ma}
\hrefCMSnoop {}{M.~Cacciari, G.~P. Salam, and G.~Soyez, ``{F}ast{J}et user
  manual'',} \textit{ Eur. Phys. J. C} \textbf{ 72} (2012) 1896,
  \href{http://dx.doi.org/10.1140/epjc/s10052-012-1896-2}{\doi{10.1140/epjc/s10052-012-1896-2}},
  \href{http://www.arXiv.org/abs/1111.6097}{\texttt{arXiv:1111.6097}}.

\bibitem{CMS-PAS-JME-16-003}
\href {https://cds.cern.ch/record/2256875}{{{CMS}} Collaboration, ``Jet
  algorithms performance in 13 {TeV} data'',} CMS Physics Analysis Summary
  CMS-PAS-JME-16-003, 2017.

\bibitem{Khachatryan:2016kdb}
\hrefCMSnoop {}{{CMS Collaboration}, ``Jet energy scale and resolution in the
  {CMS} experiment in pp collisions at 8 {TeV}'',} \textit{ JINST} \textbf{ 12}
  (2017) P02014,
  \href{http://dx.doi.org/10.1088/1748-0221/12/02/P02014}{\doi{10.1088/1748-0221/12/02/P02014}},
\href{http://www.arXiv.org/abs/1607.03663}{\texttt{arXiv:1607.03663}}.

\bibitem{Sirunyan_2018DEEPCSV}
\hrefCMSnoop {}{{CMS} Collaboration, ``Identification of heavy-flavour jets
  with the {CMS} detector in pp collisions at 13 {TeV}'',} \textit{ JINST}
  \textbf{ 13} (2018) P05011,
  \href{http://dx.doi.org/10.1088/1748-0221/13/05/p05011}{\doi{10.1088/1748-0221/13/05/p05011}},
  \href{http://www.arXiv.org/abs/1712.07158}{\texttt{arXiv:1712.07158}}.

\bibitem{Bertolini:2014bba}
\hrefCMSnoop {}{D.~Bertolini, P.~Harris, M.~Low, and N.~Tran, ``Pileup per
  particle identification'',} \textit{ JHEP} \textbf{ 10} (2014) 059,
  \href{http://dx.doi.org/10.1007/JHEP10(2014)059}{\doi{10.1007/JHEP10(2014)059}},
\href{http://www.arXiv.org/abs/1407.6013}{\texttt{arXiv:1407.6013}}.

\bibitem{Sirunyan:2019kia}
\hrefCMSnoop {}{{CMS Collaboration}, ``Performance of missing transverse
  momentum reconstruction in proton-proton collisions at {$\sqrt{s} = 13\TeV$}
  using the {CMS} detector'',} \textit{ JINST} \textbf{ 14} (2019) P07004,
  \href{http://dx.doi.org/10.1088/1748-0221/14/07/P07004}{\doi{10.1088/1748-0221/14/07/P07004}},
\href{http://www.arXiv.org/abs/1903.06078}{\texttt{arXiv:1903.06078}}.

\bibitem{Sirunyan_2018_TauIDCMS}
\hrefCMSnoop {}{{CMS Collaboration}, ``Performance of reconstruction and
  identification of {$\tau$} leptons decaying to hadrons and {$\nu_\tau$} in pp
  collisions at {$\sqrt{s}=$} 13 {TeV}'',} \textit{ JINST} \textbf{ 13} (2018)
  P10005,
  \href{http://dx.doi.org/10.1088/1748-0221/13/10/P10005}{\doi{10.1088/1748-0221/13/10/P10005}},
  \href{http://www.arXiv.org/abs/1809.02816}{\texttt{arXiv:1809.02816}}.

\bibitem{TAU-20-001}
\hrefCMSnoop {}{{CMS Collaboration}, ``Identification of hadronic tau lepton
  decays using a deep neural network'',} 2022.
  \href{http://www.arXiv.org/abs/2201.08458}{\texttt{arXiv:2201.08458}}.
  Submitted to \textit{JINST}.

\bibitem{Bianchini_2014}
\hrefCMSnoop {}{L.~Bianchini, J.~Conway, E.~K. Friis, and C.~Veelken,
  ``Reconstruction of the {H}iggs mass in {$\PH\to\PGt\PGt$} events by
  dynamical likelihood techniques'',} \textit{ J. Phys. Conf. Ser.} \textbf{
  513} (2014) 022035,
  \href{http://dx.doi.org/10.1088/1742-6596/513/2/022035}{\doi{10.1088/1742-6596/513/2/022035}}.

\bibitem{Berge:2008dr}
\hrefCMSnoop {}{S.~Berge and W.~Bernreuther, ``Determining the {$CP$} parity of
  {H}iggs bosons at the {LHC} in the {\PGt} to 1-prong decay channels'',}
  \textit{ Phys. Lett. B} \textbf{ 671} (2009) 470,
  \href{http://dx.doi.org/10.1016/j.physletb.2008.12.065}{\doi{10.1016/j.physletb.2008.12.065}},
  \href{http://www.arXiv.org/abs/0812.1910}{\texttt{arXiv:0812.1910}}.

\bibitem{Berge:2015nua}
\hrefCMSnoop {}{S.~Berge, W.~Bernreuther, and S.~Kirchner, ``Prospects of
  constraining the {H}iggs boson's {$CP$} nature in the tau decay channel at
  the {LHC}'',} \textit{ Phys. Rev. D} \textbf{ 92} (2015) 096012,
  \href{http://dx.doi.org/10.1103/PhysRevD.92.096012}{\doi{10.1103/PhysRevD.92.096012}},
  \href{http://www.arXiv.org/abs/1510.03850}{\texttt{arXiv:1510.03850}}.

\bibitem{Bower:2002zx}
\hrefCMSnoop {}{G.~R. Bower, T.~Pierzchala, Z.~W{\c{a}}s, and M.~Worek,
  ``Measuring the {H}iggs boson's parity using {$\PGt\to\PGr\PGn$}'',} \textit{
  Phys. Lett. B} \textbf{ 543} (2002) 227,
  \href{http://dx.doi.org/10.1016/S0370-2693(02)02445-0}{\doi{10.1016/S0370-2693(02)02445-0}},
  \href{http://www.arXiv.org/abs/hep-ph/0204292}{\texttt{arXiv:hep-ph/0204292}}.

\bibitem{Cherepanov:2018yqb}
\hrefCMSnoop {}{V.~Cherepanov, E.~Richter-Was, and Z.~Was, ``{M}onte {C}arlo,
  fitting and machine learning for tau leptons'',} \textit{ SciPost Phys.
  Proc.} \textbf{ 1} (2019) 018,
  \href{http://dx.doi.org/10.21468/SciPostPhysProc.1.018}{\doi{10.21468/SciPostPhysProc.1.018}},
  \href{http://www.arXiv.org/abs/1811.03969}{\texttt{arXiv:1811.03969}}.

\bibitem{Cherepanov:2018npf}
\hrefCMSnoop {}{V.~Cherepanov and A.~Zotz, ``Kinematic reconstruction of
  {$\PZ/\PH \to \tau\tau$} decay in proton-proton collisions'',} 2018.
\href{http://www.arXiv.org/abs/1805.06988}{\texttt{arXiv:1805.06988}}.

\bibitem{Jadach:1990mz}
\hrefCMSnoop {}{S.~Jadach, J.~H. Kuhn, and Z.~W{\c{a}}s, ``{TAUOLA}: A library
  of {M}onte {C}arlo programs to simulate decays of polarized {\Pgt}
  leptons'',} \textit{ Comput. Phys. Commun.} \textbf{ 64} (1990) 275,
\href{http://dx.doi.org/10.1016/0010-4655(91)90038-M}{\doi{10.1016/0010-4655(91)90038-M}}.

\bibitem{Jezabek:1991qp}
\hrefCMSnoop {}{M.~Jezabek, Z.~W{\c{a}}s, S.~Jadach, and J.~H. Kuhn, ``The
  {\Pgt} decay library {TAUOLA}, update with exact {O}(alpha) {QED} corrections
  in {$\Pgt\to\PGm(\Pe)\Pgn\Pagn$} decay modes'',} \textit{ Comput. Phys.
  Commun.} \textbf{ 70} (1992) 69,
\href{http://dx.doi.org/10.1016/0010-4655(92)90092-D}{\doi{10.1016/0010-4655(92)90092-D}}.

\bibitem{Jadach:1993hs}
\hrefCMSnoop {}{S.~Jadach, Z.~W{\c{a}}s, R.~Decker, and J.~H. Kuhn, ``The
  {\Pgt} decay library {TAUOLA}: Version 2.4'',} \textit{ Comput. Phys.
  Commun.} \textbf{ 76} (1993) 361,
\href{http://dx.doi.org/10.1016/0010-4655(93)90061-G}{\doi{10.1016/0010-4655(93)90061-G}}.

\bibitem{Asner:1999kj}
\hrefCMSnoop {}{{CLEO} Collaboration, ``Hadronic structure in the decay
  {$\Pgt\to\PGnGt\PGpm\PGpz\PGpz$} and the sign of the {\PGnGt} helicity'',}
  \textit{ Phys. Rev. D} \textbf{ 61} (2000) 012002,
  \href{http://dx.doi.org/10.1103/PhysRevD.61.012002}{\doi{10.1103/PhysRevD.61.012002}},
  \href{http://www.arXiv.org/abs/hep-ex/9902022}{\texttt{arXiv:hep-ex/9902022}}.

\bibitem{CMS-DP-2020-041}
\href {http://cds.cern.ch/record/2727092}{{CMS Collaboration}, ``Identification
  of hadronic tau decay channels using multivariate analysis ({MVA} decay
  mode)'',} CMS Detector Performance Note CMS-DP-2020-041, 2020.

\bibitem{XGBoost}
\hrefCMSnoop {}{T.~Chen and C.~Guestrin, ``{XGB}oost: A scalable tree boosting
  system'',} in \textit{ Proceedings of the 22nd ACM SIGKDD International
  Conference on Knowledge Discovery and Data Mining}, KDD '16, p.~785.
\newblock 2016.
\newblock
  \href{http://www.arXiv.org/abs/1603.02754}{\texttt{arXiv:1603.02754}}.
\newblock
  \href{http://dx.doi.org/10.1145/2939672.2939785}{\doi{10.1145/2939672.2939785}}.

\bibitem{Sirunyan_2019_Embedding}
\hrefCMSnoop {}{{{CMS}} Collaboration, ``An embedding technique to determine
  {$\tau\tau$} backgrounds in proton-proton collision data'',} \textit{ JINST}
  \textbf{ 14} (2019) P06032,
  \href{http://dx.doi.org/10.1088/1748-0221/14/06/p06032}{\doi{10.1088/1748-0221/14/06/p06032}},
  \href{http://www.arXiv.org/abs/1903.01216}{\texttt{arXiv:1903.01216}}.

\bibitem{Sirunyan:2018qio}
\hrefCMSnoop {}{{CMS Collaboration}, ``Measurement of the {$\PZ/\PGg^{*} \to
  \tau\tau$} cross section in pp collisions at {$\sqrt{s} = $} 13 {TeV} and
  validation of {$\tau$} lepton analysis techniques'',} \textit{ Eur. Phys. J.
  C} \textbf{ 78} (2018) 708,
  \href{http://dx.doi.org/10.1140/epjc/s10052-018-6146-9}{\doi{10.1140/epjc/s10052-018-6146-9}},
\href{http://www.arXiv.org/abs/1801.03535}{\texttt{arXiv:1801.03535}}.

\bibitem{Khachatryan_2011}
\hrefCMSnoop {}{{CMS Collaboration}, ``Measurements of inclusive {W} and {Z}
  cross sections in pp collisions at {$ \sqrt {s} = 7 $} {TeV}'',} \textit{
  JHEP} \textbf{ 01} (2011) 080,
  \href{http://dx.doi.org/10.1007/JHEP01(2011)080}{\doi{10.1007/JHEP01(2011)080}},
  \href{http://www.arXiv.org/abs/1012.2466}{\texttt{arXiv:1012.2466}}.

\bibitem{Khachatryan_2015}
\hrefCMSnoop {}{{CMS Collaboration}, ``Measurement of the differential cross
  section for top quark pair production in pp collisions at {$\sqrt{s} =
  8\TeV$}'',} \textit{ Eur. Phys. J. C} \textbf{ 75} (2015) 2339,
  \href{http://dx.doi.org/10.1140/epjc/s10052-015-3709-x}{\doi{10.1140/epjc/s10052-015-3709-x}},
  \href{http://www.arXiv.org/abs/1505.04480}{\texttt{arXiv:1505.04480}}.

\bibitem{deFlorian:2016spz}
\hrefCMSnoop {}{{LHC Higgs Cross Section Working Group}, ``Handbook of {LHC}
  {H}iggs cross sections: 4. deciphering the nature of the {H}iggs sector'',}
  \textit{ CERN} (2016)
  \href{http://dx.doi.org/10.23731/CYRM-2017-002}{\doi{10.23731/CYRM-2017-002}},
\href{http://www.arXiv.org/abs/1610.07922}{\texttt{arXiv:1610.07922}}.

\bibitem{CMS-PAS-LUM-17-001}
\href {https://cds.cern.ch/record/2257069}{{CMS Collaboration}, ``{CMS}
  luminosity measurements for the 2016 data-taking period'',} CMS Physics
  Analysis Summary CMS-PAS-LUM-17-001, 2017.

\bibitem{CMS-PAS-LUM-17-004}
\href {https://cds.cern.ch/record/2621960}{{CMS Collaboration}, ``{CMS}
  luminosity measurement for the 2017 data-taking period at {$\sqrt{s} =
  13\TeV$}'',} CMS Physics Analysis Summary CMS-PAS-LUM-17-004, 2018.

\bibitem{CMS-PAS-LUM-18-002}
\href {https://cds.cern.ch/record/2676164}{{CMS Collaboration}, ``{CMS}
  luminosity measurement for the 2018 data-taking period at {$\sqrt{s} =
  13\TeV$}'',} CMS Physics Analysis Summary CMS-PAS-LUM-18-002, 2019.

\bibitem{Li_2012}
\hrefCMSnoop {}{Y.~Li and F.~Petriello, ``Combining {QCD} and electroweak
  corrections to dilepton production in the framework of the {FEWZ} simulation
  code'',} \textit{ Phys. Rev. D} \textbf{ 86} (2012) 094034,
  \href{http://dx.doi.org/10.1103/physrevd.86.094034}{\doi{10.1103/physrevd.86.094034}},
  \href{http://www.arXiv.org/abs/1208.5967}{\texttt{arXiv:1208.5967}}.

\bibitem{Czakon_2014}
\hrefCMSnoop {}{M.~Czakon and A.~Mitov, ``{T}op++: A program for the
  calculation of the top-pair cross-section at hadron colliders'',} \textit{
  Comput. Phys. Commun.} \textbf{ 185} (2014) 2930,
  \href{http://dx.doi.org/10.1016/j.cpc.2014.06.021}{\doi{10.1016/j.cpc.2014.06.021}},
  \href{http://www.arXiv.org/abs/1112.5675}{\texttt{arXiv:1112.5675}}.

\bibitem{Khachatryan_2017}
\hrefCMSnoop {}{{{CMS}} Collaboration, ``Measurement of the {WZ} production
  cross section in pp collisions at {$\sqrt{s}=13$} {TeV}'',} \textit{ Phys.
  Lett. B} \textbf{ 766} (2017) 268,
  \href{http://dx.doi.org/10.1016/j.physletb.2017.01.011}{\doi{10.1016/j.physletb.2017.01.011}},
  \href{http://www.arXiv.org/abs/1607.06943}{\texttt{arXiv:1607.06943}}.

\bibitem{Sirunyan_2017}
\hrefCMSnoop {}{{CMS Collaboration}, ``Cross section measurement of
  {$t$}-channel single top quark production in pp collisions at {$\sqrt s =$}
  13 {TeV}'',} \textit{ Phys. Lett. B} \textbf{ 772} (2017) 752,
  \href{http://dx.doi.org/10.1016/j.physletb.2017.07.047}{\doi{10.1016/j.physletb.2017.07.047}},
  \href{http://www.arXiv.org/abs/1610.00678}{\texttt{arXiv:1610.00678}}.

\bibitem{BarlowBeeston}
\hrefCMSnoop {}{R.~Barlow and C.~Beeston, ``Fitting using finite {M}onte
  {C}arlo samples'',} \textit{ Comput. Phys. Commun.} \textbf{ 77} (1993) 219,
  \href{http://dx.doi.org/10.1016/0010-4655(93)90005-W}{\doi{10.1016/0010-4655(93)90005-W}}.

\bibitem{Conway:2011in}
\hrefCMSnoop {}{J.~S. Conway, ``{I}ncorporating nuisance parameters in
  likelihoods for multisource spectra'',} in \textit{ {PHYSTAT 2011}}, p.~115.
\newblock 2011.
\newblock \href{http://www.arXiv.org/abs/1103.0354}{\texttt{arXiv:1103.0354}}.
\newblock
  \href{http://dx.doi.org/10.5170/CERN-2011-006.115}{\doi{10.5170/CERN-2011-006.115}}.

\bibitem{CMS-NOTE-2011-005}
\href {https://cds.cern.ch/record/1379837}{{ATLAS and CMS Collaborations, and
  LHC Higgs Combination Group}, ``Procedure for the {LHC} {H}iggs boson search
  combination in {S}ummer 2011'',} Technical Report CMS-NOTE-2011-005,
  ATL-PHYS-PUB-2011-11, 2011.

\bibitem{Khachatryan:2014jba}
\hrefCMSnoop {}{{CMS Collaboration}, ``Precise determination of the mass of the
  {H}iggs boson and tests of compatibility of its couplings with the standard
  model predictions using proton collisions at 7 and 8 {TeV}'',} \textit{ Eur.
  Phys. J. C} \textbf{ 75} (2015) 212,
  \href{http://dx.doi.org/10.1140/epjc/s10052-015-3351-7}{\doi{10.1140/epjc/s10052-015-3351-7}},
\href{http://www.arXiv.org/abs/1412.8662}{\texttt{arXiv:1412.8662}}.

\bibitem{hepdata}
\hrefCMSnoop {}{``{HEPD}ata record for this analysis'',} 2021.
\newblock
  \href{http://dx.doi.org/10.17182/hepdata.104978}{\doi{10.17182/hepdata.104978}}.

\end{thebibliography}\endgroup
\cleardoublepage \appendix\section{The CMS Collaboration \label{app:collab}}\begin{sloppypar}\hyphenpenalty=5000\widowpenalty=500\clubpenalty=5000\vskip\cmsinstskip
\textbf{Yerevan Physics Institute, Yerevan, Armenia}\\*[0pt]
A.~Tumasyan
\vskip\cmsinstskip
\textbf{Institut f\"{u}r Hochenergiephysik, Vienna, Austria}\\*[0pt]
W.~Adam, J.W.~Andrejkovic, T.~Bergauer, S.~Chatterjee, M.~Dragicevic, A.~Escalante~Del~Valle, R.~Fr\"{u}hwirth\cmsAuthorMark{1}, M.~Jeitler\cmsAuthorMark{1}, N.~Krammer, L.~Lechner, D.~Liko, I.~Mikulec, P.~Paulitsch, F.M.~Pitters, J.~Schieck\cmsAuthorMark{1}, R.~Sch\"{o}fbeck, D.~Schwarz, S.~Templ, W.~Waltenberger, C.-E.~Wulz\cmsAuthorMark{1}
% ...
\vskip\cmsinstskip
% ...
1:  Also at TU Wien, Wien, Austria\\
% ...
\end{sloppypar}
\end{document}